\documentclass[12pt,preprint]{aastex}

\newcommand{\bdv}[1]{\mbox{\boldmath$#1$}}

\def\au{{\rm AU}} 
 
\def\kms{{\rm km}\,{\rm s}^{-1}}

\def\max{{\rm max}}
\def\rel{{\rm rel}}

\def\eff{{\rm eff}}

\def\e{{\rm E}}
\def\bpi{{\bdv\pi}}
\def\bmu{{\bdv\mu}}

\newcommand{\Spitzer}{{\em Spitzer}}

\begin{document}
\title{Criteria for Sample Selection to Maximize Planet Sensitivity and Yield from Space-Based Microlens Parallax Surveys}

\author{
Jennifer C.\ Yee\altaffilmark{1,2},
Andrew Gould\altaffilmark{3},
Charles Beichman\altaffilmark{4},
Sebastiano Calchi Novati\altaffilmark{4,5,6,7},
Sean Carey\altaffilmark{8},
B.\ Scott Gaudi\altaffilmark{3},
Calen Henderson\altaffilmark{3},
David Nataf\altaffilmark{9},
Matthew Penny\altaffilmark{3,2},
Yossi Shvartzvald\altaffilmark{10},
Wei Zhu\altaffilmark{3}
}
\altaffiltext{1}{Harvard-Smithsonian Center for Astrophysics, 60 Garden St., Cambridge, MA 02138, USA}
\altaffiltext{2}{Sagan Fellow}
\altaffiltext{3}{Department of Astronomy, Ohio State University, 140 W. 18th Ave., Columbus, OH  43210, USA}
\altaffiltext{4}{NASA Exoplanet Science Institute, MS 100-22, California Institute of Technology, Pasadena, CA 91125, USA}
\altaffiltext{5}{Dipartimento di Fisica ``E. R. Caianiello'', Universit\`a di Salerno, Via Giovanni Paolo II, 84084 Fisciano (SA),\ Italy} 
\altaffiltext{6}{Istituto Internazionale per gli Alti Studi Scientifici (IIASS), Via G. Pellegrino 19, 84019 Vietri Sul Mare (SA), Italy}
\altaffiltext{7}{Sagan Visiting Fellow}
\altaffiltext{8}{\Spitzer, Science Center, MS 220-6, California Institute of Technology,Pasadena, CA, USA}
\altaffiltext{9}{Research School of Astronomy and Astrophysics, The Australian National University, Canberra, ACT 2611, Australia}
\altaffiltext{10}{School of Physics and Astronomy, Tel-Aviv University, Tel-Aviv 69978, Israel}


\begin{abstract}

Space-based microlens parallax measurements are a powerful tool for
understanding planet populations, especially their distribution
throughout the Galaxy. However, if space-based observations of the
microlensing events must be specifically targeted, it is crucial that
microlensing events enter the parallax sample without reference to the
known presence or absence of planets. Hence, it is vital to define
objective criteria for selecting events where possible and to
carefully consider and minimize the selection biases where not
possible so that the final sample represents a controlled experiment.

We present objective criteria for initiating observations and
determining their cadence for a subset of events, and we define
procedures for isolating subjective decision making from information
about detected planets for the remainder of events.  We also define
procedures to resolve conflicts between subjective and objective
selections.  These procedures maximize planet sensitivity of the
sample as a whole by allowing for planet detections even if they occur
before satellite observations for objectively-selected events and by
helping to trigger fruitful follow-up observations for
subjectively-chosen events.  This paper represents our public
commitment to these procedures, which is a necessary component of
enforcing objectivity on the experimental protocol.

\end{abstract}


\section{{Introduction}
\label{sec:intro}}

\subsection{Measuring the Distances to Microlensing Planets}

While more than 6000 planets (and strong planetary candidates) have
been found within about 1 kpc of the Sun (the great majority
discovered via the transit and radial velocity techniques), there are
only a handful of confirmed planets with known distances that are
greater than 4 kpc and only one confirmed planet in the Galactic bulge
\citep{mb11293B}.  All of these distant planets were found using
gravitational microlensing, and in most cases the distances were
determined using the ``microlens parallax'' technique \citep{gould92}.
Microlensing would therefore appear to be the most natural method to
measure the Galactic distribution of planets, i.e., to determine
planet frequency as a function of Galactic environment.  Such a
measurement would provide important constraints on planet formation
theories.  For example, \citet{thompson13} has suggested that
gas-giant formation may have been inhibited in the Galactic bulge due
to the high intensity of ambient radiation during the main epoch of
star formation.

However, while roughly half of the $\sim30$ published microlensing
planets have measured distances, this sample is heavily biased toward
nearby systems.  The reasons for this are well understood and are
closely related to the general biases in astronomy toward nearby
objects.  First, nearby lenses have larger lens-source trigonometric
parallaxes, $\pi_\rel = \au(D_L^{-1}-D_S^{-1})$, which gives rise to
larger microlens parallaxes
\begin{equation}
\bpi_\e \equiv {\pi_\rel\over\theta_\e}\,{\bmu\over\mu};
\qquad
\theta_\e^2 = \kappa M\pi_\rel,
\qquad \kappa \equiv {4 GM\over c^2\au}\simeq 8.1\,{{\rm mas}\over M_\odot},
\label{eqn:pie}
\end{equation}
where $\bmu$ is the lens-source relative proper motion (in either the
heliocentric or geocentric frame), $\theta_\e$ is the angular Einstein
radius, and $M$ is the lens mass.  As explained in some detail by
\citet{gouldhorne}, the magnitude of $\bpi_\e$ quantifies the
amplitude of the parallax distortion on the microlens light curve, so
that all other things being equal, larger $\pi_\e$ implies easier
detection.  The most common method for measuring microlens parallax
has been to observe the effect of Earth's acceleration on the light
curve (so-called orbital parallax).  However, for typical Einstein
timescales $t_\e\sim 20\,$day, this effect is quite modest.  This
means that in addition to nearby lenses and low mass lenses, one is
biased toward abnormally long duration events.  It is difficult
(though probably not impossible) to quantify these biases, but the
main problem is that due to these biases, there are simply no
microlens planets in the Galactic bulge with measured microlens
parallaxes.  Indeed, the one confirmed Bulge planet had its distance
measured by other means.

This brings us to the other method of measuring lens distances: direct
detection of the lens.  The main difficulty is that the lens is
superposed on a (usually) substantially brighter source, and remains
so for typically a decade or more after the event.  If the lens is
sufficiently bright, then it is possible to directly detect it by
measuring the combined source and lens light using high-resolution
imaging (adaptive optics or {\it Hubble Space Telescope (HST)}) and
subtracting out the source contribution, which is known from the light
curve model.  This, in fact, is how the distance to the only planet
known to be in the Galactic bulge was measured
\citep[MOA-2011-BLG-293Lb; ][]{mb11293B}.  At the present time, this
method is primarily limited to lenses that are at least 15\% as bright
as the source: otherwise the excess light due to the lens cannot be
reliably detected.  Hence, it is biased toward luminous (i.e.,
massive) and nearby lenses.

The alternative is to wait until the source and lens separate due to
their relative proper motions (typically a few mas yr$^{-1}$) and can
be individually resolved. Again, this method is more easily applied to
brighter lenses and with current facilities, one must wait $\sim10$ yr
for the source and lens to separate sufficiently.  When the next
generation of 30 m telescopes are available, it will be applicable to
much fainter lenses because these will separate sufficiently from the
sources to be resolved within a few years due to their relative proper
motions \citep{alcock01,gould14,ob05169a,ob05169b,henderson15}.

Therefore, the only path at present to routinely measure the distances
to lenses (especially faint lenses), and hence to measure the Galactic
distribution of planets, is via space-based microlens parallaxes.  In
this approach, one observes a microlensing event simultaneously from
Earth and from a satellite in solar orbit, and derives $\bpi_\e$ from
the difference in the two light curves \citep{refsdal66}.  There are
some challenges to this method (over and above the problem of gaining
routine access to such a satellite).  First, the results are subject
to a four-fold degeneracy in $\bpi_\e$, including a two-fold
degeneracy in $\pi_\e$.  However, \citet{21event} showed that it is
possible in practice to break this degeneracy in the great majority of
cases.  Second, $\pi_\e$ does not by itself yield distances and
masses.  Rather this requires knowledge of $\theta_\e$,
\begin{equation}
\pi_\rel = \theta_\e \pi_\e,
\qquad
M = {\theta_\e\over \kappa \pi_\e},
\label{eqn:massidst}
\end{equation}
and of the source parallax $\pi_S$ ($\pi_L = \pi_\rel + \pi_S$),
although the latter is usually known quite adequately.  Fortunately,
$\theta_\e$ is usually measured for planetary events because the
normalized source size $\rho\equiv\theta_*/\theta_\e$ can usually be
measured from the source crossing of the planetary caustic, while the
angular source size $\theta_*$ is almost always known from its color
and magnitude.  Moreover, even for non-planetary (and non-binary)
events, which generally lack such crossings, the lens distance (and so
mass) can usually be estimated quite well from the measured $\bpi_\e$
and kinematic arguments \citep{21event}.  Finally, for the case that
the source proper motion can be measured, this estimate becomes even
more accurate \citep{ob140939}.

Hence, as shown by \citet{21event}, one can obtain an accurate
estimate of the cumulative distribution of lens distances from a given
sample, and can in principle compare this to the cumulative distance
distribution of detected planets.

\subsection{\Spitzer\, and the Galactic Distribution of Planets}

To determine the Galactic distribution of planets, however, the
detected planets must be compared to the underlying distribution of
planet sensitivities, not simply of events.  \citet{21event} did not
attempt to do this because there was only one planet in their sample
\citep{ob140124}, making a meaningful comparison impossible.  The
small number of planet detections was rooted in the nature of the
observing campaign, which was a 100-hr ``pilot project'' to determine
the feasibility of making such microlens parallax measurements using
{\em Spitzer}.  Thus, the {\em Spitzer} observations were limited to
the subset of events judged most likely to yield $\bpi_\e$, and no
special effort was made to find planets within these events via, for
example, intensive follow-up observations.

\citet{21event} argued, nevertheless, that it would be possible to
estimate the cumulative distribution of sensitivities, simply by
measuring the sensitivity of each event in the standard fashion
\citep{rhie00,gaudisackett,gaudi02} 
and multiplying these sensitivities by the distance
distributions in their Figure 3, even though the selection function
of the events was unknown (and probably unknowable).  
This argument rested critically
on the fact that the events were monitored from the ground and
chosen for {\em Spitzer} observations without regard to the presence
of absence of planets.
This is a very similar argument to the one
made by \citet{gould10} in the first study to derive planet frequencies
from microlensing planet detections.  \citet{21event} further argued that their
sample could be concatenated with future space-based samples, regardless
of whether these were carried out using {\em Spitzer} or other satellites
such as {\it Kepler}, and regardless of whether the selection function
was the same or different, known or unknown.  The only proviso was that,
as with  the \citet{gould10} and \citet{21event} samples, the events 
were monitored without regard to the presence or absence of planets.

\subsection{This Paper}

\begin{deluxetable}{ll}
\rotate
\tablecaption{Glossary of terms\label{tab:glossary}}
\tablehead{\colhead{Term} & \colhead{Definition}
}
\startdata
\sidehead{General:}
planet sensitivity \dotfill 
& The range of planets (as a function of $s$ and $q$) that could \\
& hypothetically lead to measureable signals in the light curve. \\
& Also called ``detection efficiency.''\\
\hline
\sidehead{Characterization of an event: }
``good'' event \dotfill
& An event with significant sensitivity to planets and a high \\
& likelihood of yielding a parallax measurement.\\
``bad'' event \dotfill 
& An event with poor sensitivity to planets or a low likelihood \\
& of yielding a parallax measurement.\\
\hline
\sidehead{General light curve terminology: }
the ``peak'' of an event \dotfill 
& The part of the light curve encompassing $t_0$\\
the ``wings'' of an event \dotfill 
& The part of the light curve with $|t-t_0|\gtrsim t_\e$\\
``rising'' \dotfill 
& An event is rising if $t<t_0$. Likewise, the ``rising'' part of\\
&  the light curve refers to the part of the light curve with $t<t_0$.\\
``falling'' \dotfill 
& An event is falling if $t>t_0$. Likewise, the ``falling'' part \\
& of the light curve refers to the part of the light curve with $t>t_0$.\\
``over'' \dotfill 
& An event is over if it has returned to its baseline (unmagnified) \\
&  magnitude.\\
\enddata
\end{deluxetable}

The goal of the \Spitzer\, microlensing parallax program (and indeed
any space-based parallax program) is to create a sample of events with
well-measured parallaxes. If these events are observed by the
satellite without regard to whether or not they have planets, the
final sample can be used to determine the Galactic distribution of
planets, e.g., by comparing the frequency of planets in the Galacitic
bulge with the frequency of planets in the disk. Hence, achieving this
scientific goal has three primary considerations. First and foremost,
the decision to select an event for \Spitzer\, observations must be
independent of any knowledge of the presence or absence of a
planet\footnote{While it is more natural to think about the presence
  of planet creating bias in the sample, the absence of planets is
  also important. For example, if an event does not show evidence of a
  planet, it could be excluded from selection for parallax
  measurements because it is ``uninteresting.'' Hence, the absence of
  planets has similar potential to create bias in the final sample but
  in the opposite direction as the presence of planets.}.  Second,
these observations must lead to a measurable parallax. Finally,
maximizing the constraints on the Galactic distribution of planets
requires maximizing not only the number of planets detected but also
the range of planets that {\it could} be detected (i.e. the planet
sensitivity), since the detection efficiency is a crucial component to
any measurement of the planet occurrence rate.

The primary goal of the present paper is to determine a strategy to
monitor events with \Spitzer\, to ensure the final sample of events
with parallaxes is monitored without reference to the presence or
absence of planets, because this is the property of the sample that is
most difficult to control.  At the same time, this strategy is driven
by the additional goals of maximizing both the planet sensitivity of
the monitored events and the likelihood of measuring parallaxes. By
defining this strategy in advance of the observations, we can create a
sample of events with measured parallaxes with maximal leverage for
measuring the occurrence rate of planets as a function of Galactic
distance.

We begin, in Section \ref{sec:objsub}, with a general discussion of
how events may be selected, either objectively or subjectively, and
how that selection affects the resulting planet sensitivities of those
events. Since much of that discussion is guided by planet sensitivity
and the practical considerations of the \Spitzer\, campaign, the
reader may also wish to refer to Sections \ref{sec:planet-sens} and
\ref{sec:spitzobs}. Sections
\ref{sec:ingredients}--\ref{sec:parallaxprob} cover the various
ingredients necessary to define criteria for selecting events, namely
the planet sensitivity and the probability of measuring parallax. Then
in Section \ref{sec:objcrit}, we formally define objective criteria
for the \Spitzer\, campaign to select events and also determine their
observing cadences. Section \ref{sec:subchoice} then discusses
specific guidelines for subjectively choosing events for this
campaign, and Section \ref{sec:realloc} specifies how the available
observations will be distributed amongst the targets. Finally, we give
a brief summary in Section \ref{sec:conclude}. We have provided a
glossary of terms in Table \ref{tab:glossary} to clarify some
qualitative statements we may use and also colloquialisms that have
arisen in microlensing.

{\section{Objective vs.\ Subjective Selection Criteria}
\label{sec:objsub}}

\begin{deluxetable}{lll}
\rotate
\tablecaption{Definitions of Relevant Times\label{tab:definitions}}
\tablehead{\colhead{Variable} & \colhead{Quantitative (approx.)} & \colhead{Qualitative}
}
\startdata
$j$ \dotfill                  & 1--7 \dotfill & Week of observations (beginning Thursday and\\
&& ending Wednesday)\\   

$t_{\rm first}$ \dotfill     & HJD$^{\prime}$7182.5 & First date of {\em Spitzer} observations (set by the \\
&= 9 June 2015  \dotfill &allotted observing window or Sun-angle constraints)\\

$t_{\rm fin}$ \dotfill       & HJD$^{\prime}$7222.5& Last date of {\em Spitzer} observations (set by the\\
&= 19 July 2015  \dotfill &allotted observing window or Sun-angle constraints)\\
$t_{j,\rm dec}$ \dotfill      & Monday UT 9:00  \dotfill& Time when {\em Spitzer} observations are finalized for Week $j$\\
$t_{j,\rm next}$  \dotfill    & $t_{j,\rm dec}+3$ days = Thursday  \dotfill& Time of first {\em Spitzer} observations for Week $j$\\
$t_{\rm sel}$  \dotfill  & $t_{j-1,\rm dec}<t_{\rm sel}\leq t_{j, \rm dec}$ \dotfill & Time when event is selected for {\em Spitzer} observations\\
$t_{\rm com}$  \dotfill& $\geq t_{\rm sel}$  \dotfill& Time of public commitment to an event\\
$t_{\rm stop}$  \dotfill& $\leq t_{\rm fin}$  \dotfill& Time when {\em Spitzer} observations cease\\
\enddata
\tablecomments{Although these terms are defined relative to the
  {\em Spitzer} campaign that is described in this paper, they are easily
  generalized for any targeted microlens parallax program.}
\end{deluxetable}

\begin{deluxetable}{lll}
\rotate
\tablecaption{Ways to Select an Event\label{tab:selection}}
\tablehead{\colhead{Type} & \colhead{This event...} & \colhead{Planets and sensitivity from...}
}
\startdata
Objective  & 
    Meets pre-defined criteria. & 
    The entire event may be counted. \\
\hline
Subjective, & 
    Does not meet pre-defined criteria,& 
    The selection date ($t_{\rm sel}=t_{\rm com}$) \\
immediate &  but is selected anyway. &onwards may be counted.\\
\hline
Subjective, & 
    Has unconstrained fits and may or & 
    The selection date ($t_{\rm sel}=t_{\rm com}$) \\
conditional  & may not be ``good''. It is selected &onwards may be counted\\
  & anyway but with specific conditions \\
  & (defining event failure) for halting \\
  & observations. \\
\hline
Subjective, & 
    Has unconstrained fits and may or & 
    The committment date \\
secret & may not be ``good''. It is selected & ($t_{\rm com}>t_{\rm sel}$) onwards may be counted.\\
  & anyway but may be dropped at any time \\
  & until the event is publicly committed to.\\
\enddata

\tablecomments{There are four ways an event can be selected, with
  three distinct modes of subjective selection. These simply reflect
  ways in which events may enter the sample. Any events that
  ultimately do not yield a parallax measurement will be removed from
  the sample, regardless of how they are selected.}

\end{deluxetable}

There are many choices that must be made with respect to {\em Spitzer}
observations of any individual event.  One must decide when to begin making
such observations, when to commit to the 
target\footnote{The distinction between deciding to observe a target
and committing to a target is discussed in Section~\ref{sec:sub}.}, 
with what initial
cadence, whether and when to change this cadence, and whether and when
to halt the observations. 
This entire chain must be carefully established in order
to ensure the fundamental requirement that the observational sequence 
be indifferent to the
presence or absence of planets.
Table \ref{tab:definitions} gives a brief overview of the relevant
decision points, and Section \ref{sec:spitzobs} discusses how the
specifics of \Spitzer\, operations set the quantitative definitions.

The starting point is the choice to begin monitoring an individual
event, i.e., ``triggering'' observations.  
This choice can be made either because the event meets some
objective criteria (in which case the ``choice'' is automatic) or
according to some subjective criteria of the team organizing the
observations. Table \ref{tab:selection} summarizes the various channels through which observations may be triggered. However,
all other decisions about the monitoring are heavily
influenced by the first dichotomy (objective vs. subjective), so we divide the discussion
according to it.  As we will describe, optimal event selection
requires a combination of objective and subjective selection.
Because this is so, one must also decide what to do in advance
if the objective and subjective selection procedures collide.  That is, what
should be done if an event is selected subjectively, but later
meets the objective criteria for selection.

We discuss the architecture of the selection procedure before discussing
the criteria themselves 
because the architecture is both non-trivial and logically independent 
of the criteria. Within the framework of this discussion, one must 
keep in mind
that the overall goal is to maximize the sensitivity of the experiment
to planets and that planet sensitivity rests primarily on ground-based
observations (see Section~\ref{sec:planet-sens}). 
At the same time, after an event is selected, its entry into the final
sample to measure the Galactic distribution of planets requires that
its parallax is measured, which depends primarily on {\em Spitzer}
observations. 

In the following sections, we occasionally give examples to
illustrate the points under discussion. For these examples, it may be
helpful to keep in mind that some of observables that affect planet
sensitivity and parallax measurements include the time of the event
peak $t_0$, the magnification of the event (larger is better), and the
magnitude of the event as seen from the ground or from \Spitzer\,
(brighter is better). These observables and their relationship to
planet sensitivity and parallax are discussed in detail in Sections
\ref{sec:planet-sens} and \ref{sec:parallaxprob}. The final criteria
are given in Section \ref{sec:objcrit}.

{\subsection{Objectively Chosen Events}
\label{sec:obj}}

\subsubsection{Objective Selection}

The great advantage of choosing events by objective criteria is that
any planet that is discovered during an event that is so chosen can be
included in the sample, and similarly, the planet sensitivity of the
event over its entire duration can be included in the analysis as
well.  For example, suppose an event is announced by a survey group on
May 1 but \Spitzer\, observations cannot begin until June 8. The event
undergoes a planetary deviation on May 15, peaks on May 28, and on
June 3 is scheduled for \Spitzer\, observations beginning June 8
because it is found to meet previously chosen objective criteria.
Then the planet can be included in the sample, even though it was
discovered before the {\em Spitzer} observations began, and even
before it was known that it would eventually satisfy the objective
conditions that triggered observations.

By contrast, in the absence of such criteria, the event could have
been selected for {\em Spitzer} observations subjectively.  In that
case, neither the previously discovered planet, nor the planet
sensitivity from the the entire pre-decision period, could be included
in the analysis.  Otherwise, the presence or absence of the planet
could influence the event's ``selection'' (i.e., inclusion in the
final sample with measured parallaxes; Section \ref{sec:measurement} discusses
what is meant by a ``measured parallax'').

\subsubsection{Objective Cadences}

A large fraction of objectively chosen events will be similar
to the hypothetical one described above in that most of their planet
sensitivity will be past at the time that the {\em Spitzer} observations
begin.  Therefore, it is absolutely essential that the cadence be
chosen objectively as well. 
In order to enter the sample, the event must have a measured parallax. If the cadence is not chosen objectively,
events with planets could
receive extra observations to help ensure they have measured
parallaxes.
We will discuss specific algorithms to make this choice in 
Section~\ref{sec:objcrit}.

Finally, we will just mention that to avoid wasting observations, there
must also be a mechanism for halting this objectively-determined {\em Spitzer} 
observation schedule when these observations
are no longer useful.  However, the only
permitted reasons for doing so are that the microlens parallax has
been measured or that the event (as seen from {\em Spitzer}) has
already returned to baseline (i.e., is now essentially unmagnified), so that no further improvement is possible.

{\subsection{Subjectively Chosen Events}
\label{sec:sub}}

\begin{figure}
\includegraphics[trim=1in 2in 0in 2.5in]{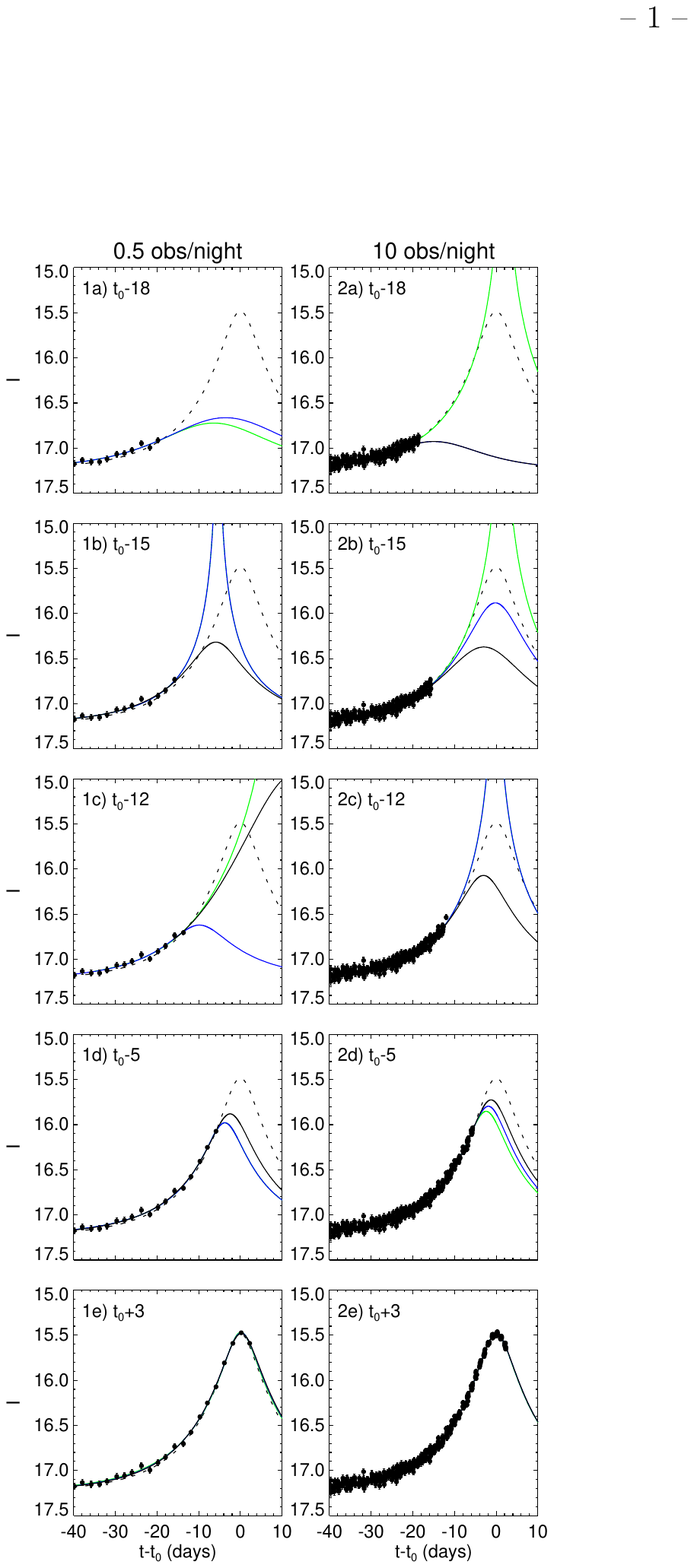}
\caption{Example fits to simulated event data. The dashed line shows
  the underlying point lens ($u_0=0.2$, $t_\e=20$ days) from which the
  data were generated with Gaussian errors. The right-hand panels show
  data sampled 10 times/night, every night, and the left-hand panels
  show the same data, but thinned to 1 observation every 2 nights. Starting
  from the top, the rows show the approximate 1-$\sigma$ range of fits
  (solid lines) to the data at $t=18,15,12,5$ days before the peak and
  at $t=3$ days after the peak. These figures show that the fits
  to the high-cadence data (10 obs/night) are much better at capturing
  the true, underlying behavior of the events, whereas fits to the
  sparse data (0.5 obs/night) often give misleading fits that do not
  encompass the true event. \label{fig:lcs}}
\end{figure}

\subsubsection{The Need for Subjective Selection}

At first glance, the advantage conferred by the objective approach appears to be
so great that one might wonder why one would consider the subjective
approach at all.  
The main problem is it is impossible to define objective criteria
loosely enough to capture all events of interest without at the same
time introducing a large number of events either with poor planet
sensitivity or a low probability of yielding a parallax measurement
(``bad'' events).  Hence, the objective criteria must be strictly
defined so that all of the selected events are both highly sensitive to
planets and have a high likelihood of yielding a parallax (``good''
events). Otherwise, a large amount of observing time will be wasted on
events of little value.

A second issue is whether or not an event chosen objectively will
yield a parallax measurement. The central difficulty is that the
event's objectively-chosen {\em Spitzer} observational sequence must
yield a parallax measurement. It can be quite difficult to choose such
observational sequences based solely on objective criteria, or to
determine which events might be worth the observational effort to
obtain parallaxes, or even to determine which might yield parallax
measurements with any sort of effort.  These difficulties can all be
more effectively addressed by subjectively choosing the event, in
which case one can also choose a cadence (or cadence algorithm) that
is individually tailored to that event.  For example, suppose that it
is known that an event will meet objective criteria in 2 weeks, but
this will allow for only 1 week of \Spitzer\, observations. If we wait
to start observations until this date, we risk the possibility that 1
week of data will be insufficient to measure a parallax, in which case
the entire event and its sensitivity is lost. In contrast, we could
select the event subjectively now to get 3 weeks of \Spitzer\,
observations and vastly improve the probability that those
observations will yield a parallax.

Hence, there are two reasons that events might be chosen
subjectively. First, because the objective criteria cannot capture all
events of interest. Secondly, because an earlier subjective trigger
may make the difference between measuring a parallax or not. As we
discuss in the next section, not much planet sensitivity is lost by
subjectively choosing events. We will also discuss in Sections
\ref{sec:collide} and \ref{sec:reversions} the resolution of
conflicts for events that may be selected both subjectively and
objectively.

\subsubsection{Subjective Selection}

If, for whatever reason, an event fails to meet the objective
selection criteria, the team may decide to observe it anyway.
The reasons one might want to do this are discussed in some detail
below, but first we focus on the consequences of this
decision, which leads to three types of subjective selection as defined in Table \ref{tab:selection}.

Actually, the key decision is not whether to begin {\em Spitzer}
observations but when to (publicly) commit to a schedule of
{\em Spitzer} observations.  Once such a decision is made, it must
be accompanied by a public commitment of an observational sequence
(or to an objective algorithm for determining that sequence).  Otherwise,
one could choose events without any knowledge of whether they would
later show planets, discover that they do indeed host planets, and then
be biased to observe them more frequently in order to preferentially 
increase the probability that their parallax will be measured, thus
placing them in the sample for measuring the Galactic distribution of planets.

In the case of subjective decisions, all planets that show 
up after the public commitment date ($t_{\rm com}$),
as well as all sensitivity to planets after this date, would be
kept in the analysis, while all planets and planet sensitivity from
before this date must be excluded.

We illustrate the need to separate the {\it decision} to observe from the
{\it commitment} to observe with two examples.  

First, suppose that at a given {\em Spitzer} decision time (typically
$t_{\rm sel}=$Monday, see Section~\ref{sec:spitzobs} for details of
the logistical constraints on {\em Spitzer} observations) an event
that has not yet peaked has an ambiguous future, with some chance that
it will rise in magnification sufficiently to have good sensitivity to
planets and to enable a viable parallax measurement (e.g., panel 2a of
Figure \ref{fig:lcs}).  However, as of Monday, this cannot be
established with any confidence, although ground-based data are likely
to resolve this ambiguity with a few days.  Based on this assessment,
the team decides to observe this event once per day during the week
beginning at the next upload three days hence, i.e., $t_{j,\rm
  next}=$Thursday.  Even though the future of the event is uncertain,
preemptive observations (rather than waiting for the next upload
cycle, one week hence) could make the difference between a good
parallax measurement and a meaningless upper limit.  This is
especially true since the {\em Spitzer} observations can only be
updated once per week.  Then on Sunday, ground-based data show that
this event has risen sufficiently that its future behavior can be
predicted well enough to determine that it is an interesting target
(e.g. panel 2c of Figure \ref{fig:lcs} as compared to panel 2a).  The
team then announces that it is committed to monitoring this event and
also announces its chosen cadence (or objective procedure for
determining such cadence).  All planets and planet sensitivity from
$t_{\rm com}=$Sunday forward can then be included in the analysis.
Similarly, only {\em Spitzer} data beginning on Sunday can be used
in the initial parallax measurement measurement that determines
whether the event enters the sample.  Even though the
Thursday--Saturday observations cannot be used, this preemptive
decision to observe has resulted in extra observations (Sun--Wed) that
can be used to improve the parallax measurement as compared to waiting
until the following Thursday to begin {\em Spitzer} observations.
With regard to the specific role of {\em Spitzer} data in planet
sensitivity and discovery: if the team has seen the first few days of
{\em Spitzer} data prior to the Sunday commitment, then these cannot
be included, but if they have not yet seen them, these can be (since
in this case they would not have influenced the decision).

For a second example, consider the same case as above but with the
event peaking at low magnification, and hence having both low sensitivity
to planets and low likelihood of a measurable parallax (e.g., the data instead follow the solid black line in panel 2a of Figure \ref{fig:lcs}).
The team then decides not to continue monitoring this
event. Because the team never committed to the event, they have no obligation
to continue monitoring it, and so it is entirely dropped from the sample.
By the same token, any planets discovered
from this event cannot be included in the analysis.
In contrast, if the decision and commitment were the same process, 
in order to avoid bias, the team would be required to continue
monitoring an event that it recognized as worthless.

In the first example given above, the public commitment to observe
the event was made after the {\em Spitzer} observation sequence was uploaded to
the spacecraft and indeed after those observations began ($t_{\rm com}>t_{j,\rm next}>t_{\rm sel}$). This is the situation described as ``Subjective, secret'' in Table \ref{tab:selection}. However, the
same principles apply to, for example, an event that is newly recognized
a few days before the upload decision and is already recognized to
be promising (e.g. panels 2c or 2d of Figure \ref{fig:lcs}).  The team could publicly commit to observing this
event immediately with specified cadence (i.e. ``Subjective, immediate'' in Table \ref{tab:selection}), 
and then all planets and planet sensitivity from that date forward
would be included (provided the parallax was measured well enough
to put the event in the sample). 

Finally, in the first example, the team might publicly commit
immediately to observe the ambiguous event, but with explicitly stated
criteria for halting such observations (Table \ref{tab:selection}, ``Subjective, conditional'').  For example, it might
specify that observations would be discontinued if the event failed to
reach some specified magnification before the time of the next upload.
In this case, whether or not the observations were continued, the
event might have sufficient {\em Spitzer} data to measure the parallax
and enter the sample.  If so, all planets and planet sensitivity from
after $t_{\rm sel}=t_{\rm com}$ could enter the sample.

At first sight, ``Subjective, conditional'' seems clearly superior to ``Subjective, secret'' 
because it enables inclusion of more planets
and more planet sensitivity.  However, this is not always the case.
In fact, it often happens that the uncertainty in predicting rising
events includes not only to their time of peak and peak magnification,
but extends even to when this knowledge will be reliably available. 
For example, consider the evolution of the fits to the data in column
1 of Figure \ref{fig:lcs}, which illustrates how the fits may change
dramatically
due to small fluctuations in the data and may not capture the true,
underlying behavior of the event.
If the team cannot reliably predict the future course of the event,
it may not be possible to correctly pre-define criteria for halting
observations.  This creates the risk of being committed to observing
many bad events or being forced to halt observations of an event 
that turns out to be good, but with different parameters than
initially supposed.

Thus, as summarized in Table \ref{tab:selection}, subjective decisions can take a considerable variety of forms.
The only constraint is that they must be constructed to avoid
the possibility that the presence or absence of planets detected
after $t_{\rm com}$ will influence the cadence of observations.

\subsubsection{Subjective Cadences}

In contrast to objectively chosen events, the cadences for subjectively
chosen events can be chosen by the team.  However, they must be
fully specified at the time of the commitment to observations ($t_{\rm com}$).  As with objectively
chosen events, after $t_{\rm com}$ the only permitted reasons for halting the scheduled {\em Spitzer}
observations
are that the parallax has already been measured or the
event has returned to baseline (as seen from {\em Spitzer}).

{\subsection{Collisions: Subjectively Chosen Events that Meet Objective Criteria}
\label{sec:collide}}

An event that has been subjectively chosen may, at a later date,
meet the objective criteria.  In this case, the objective selection
of the event {\it must} take precedence. 
Otherwise, there is no point in having objective criteria since they
could always be subjectively overridden. The objective criteria will specify
an objectively-determined
cadence of observations. The more frequent of the subjective or objective cadences will take precedence for future observations. However, 
only the observations taken after the objective selection and at the objective cadence can
be used to determine whether the parallax measurement is adequate
to enter the sample.  If it does enter, then all planets that are
detected and all planet sensitivity (including from before the
commitment to the event, $t_{\rm com}$) enter the analysis. 
The next section will clarify the reason for maintaining a higher, but
subjective, cadence even if only a fraction of the data can be used
for the objectively-measured parallax.

Note that for these cases
all planets discovered in {\em Spitzer} data (as well as
all planet sensitivity from these data) should be included in the
analysis, regardless of whether these data were taken before or after
or in response to
the commitment to the subjective selection of the event. 
The only exception would be if the planet were detected purely in the
\Spitzer\, data {\it and} subsequent \Spitzer\, observations were
increased because of it (see Section \ref{sec:characterization}).

{\subsection{Reversions from Objective to Subjective}
\label{sec:reversions}}

If an event has 1) been chosen subjectively, 2) subsequently satisfied
the objective criteria and thus triggered a conversion to being 
objectively-chosen, 3) fails to yield a parallax based on the
objective portion of the {\em Spitzer} light curve, and 4) does
yield a parallax based on all, post-commitment {\em Spitzer} data, 
then it automatically reverts to subjective status.  In this case,
only planets and planet sensitivity from the post-commitment part
of the light curve can be included in the analysis.

The basic reason is that none of the decisions made after this
commitment depended in any way on the presence or absence of a 
planet (other than possible planets before this date, which must
be excluded from the calculation of planet occurrence).

{\subsection{Planet-Characterization Observations}
\label{sec:characterization}}

As discussed above, whether the decision to observe an event with {\em Spitzer}
is objective or subjective, the cadence must be chosen without
reference to the presence or absence of a planet.  For the objectively-chosen
events, this cadence must be determined objectively by pre-determined
criteria.  For subjectively chosen events, they must be announced at
the times of the decision, i.e., prior to the discovery of any planets.

However, once a planet is discovered, it can be important to increase
the pace of observations from the ground and/or {\em Spitzer}
in order to improve the {\it characterization} of
the planet.  In this case, one can increase the observational cadence,
but only the observational data that would have been taken under the
pre-determined schedule can be used to assess the {\it detectability} of
planets and the measurability of the parallax.  See, for example,
\citet{mb11293}.  Application of this rule is straightforward for
{\em Spitzer} observations, which is the main focus in this section,
because, as we have specified, 
the observational cadence is in fact pre-determined.  The
situation is more complicated for ground-based observations of the
same event, to which we now turn.

{\subsection{Ground-based Follow-up Observations}
\label{sec:gbfollow}}

The majority of microlensing planets published to date were discovered
by a combination of microlensing surveys that find large numbers of
events, mostly with a low-to-moderate cadence of observations, and
follow-up surveys that target individual events for more intensive
monitoring in order to enhance the discovery and characterization
of planets.  In fact, survey groups sometimes go into ``follow-up mode''
by increasing the cadence in the survey field that contains a
particularly interesting event \citep{mb11293} or
even re-centering an existing field to incorporate 
a particular event \citep{mb13220}.
For present purposes, observations by specialized follow-up groups and
survey teams in ``follow-up mode'' are equally considered as ``follow up''.
The only exception would be follow-up observations that are determined 
by purely objective criteria.  

Follow-up observations must be evaluated with respect
to two questions: first, how do they impact planet sensitivity 
and planet detection, and second how do they impact the measurement
of microlens parallax (and so entry into the sample).

\subsubsection{Follow-Up and Planet Sensitivity/Detection \label{sec:followupsens}}

As with survey observations, follow-up observations contribute to
planet sensitivity and detection for the entire duration of
objectively chosen events and for all post-commitment observations
of subjectively chosen events (which may be a time-span much longer than the window for \Spitzer\, observations).  Indeed, while the fundamental point
of announcing subjective choices of events for {\em Spitzer} observations 
(i.e., the commitment to observe an event)
is to establish a record of what planet sensitivity can be included,
a major secondary goal is to encourage observations of these events,
particularly those that are not well covered by the surveys,
(see Section~\ref{sec:ground} for a general outline of standard survey
strategies).  Such subjective
announcements automatically have the effect of encouraging
follow-up observations because prior to such
announcement, the planet detections can enter only if the event
has already satisfied, or ultimately proves to satisfy, objective
criteria.

The only question is whether changes in the adopted cadence of 
follow-up observations due to the perceived presence of planets
influences the detectability of the planets.  This can happen
in principle if the planet generates an observed perturbation 
(in either survey or follow-up data) that is strong enough to trigger 
interest but not, by itself, sufficient to confirm the presence of the
planet.  In this case, follow-up observations aimed at characterizing
the planet can make the difference between it being undetectable
and being detectable.

This issue is not particular to parallax experiments: it pertains
to any experiment that aims to make a statistical statement about
planets using microlensing follow-up data.  For example, \citet{gould10}
noted that two of their six detected planets occurred in an event that
showed an early (and in itself, not comprehensible) perturbation that
ultimately proved to be due to a planet.  This early perturbation did trigger
additional observations, but these subsided over the next few days.
Observations only intensified again when the event approached 
high magnification, which was the standard trigger for high-cadence
follow-up observations.

In brief, these issues arise in a minority of planetary events,
and usually can be resolved based on records of the decision
making.  While it is possible that there may be unresolved cases
in the future, the importance of characterizing planets is too great
to allow this possibility to interfere with aggressive follow-up
response to the tentative detection of planets.

\subsubsection{Follow-Up and Parallax}

The other aspect that must be considered is the role of follow-up
observations in measuring microlens parallaxes and so in putting
individual events into the final sample used to measure the Galactic distribution of planets.  There are two relatively
distinct ways that this can happen.  First, the planetary-induced features
in the light curve may substantially increase the precision of the
parallax measurement from {\it ground-based observations alone} 
and so make the difference between whether it is included in or excluded from the sample.  Second, 
the follow-up observations may improve the precision of the non-planetary 
light curve parameters $(t_0,u_0,t_\e)_\oplus$ (i.e., the time of maximum, the impact parameter, and the Einstein timescale) and thus improve the precision
of the parallax measurement that comes from the comparison with {\em Spitzer}
data.  We discuss these in turn.

\citet{angould01} argued that events with three peaks (features due to
caustics induced by a companion) would gain significantly improved
parallax measurements relative to otherwise similar point-lens events.
However, in the intervening years, almost nothing has been done to
investigate the role of the perturbations in the parallax measurements
for detected planets.
 For example, early modeling showed that the immediate
post-peak light curve of MOA-2009-BLG-266 yielded a surprisingly good
parallax measurement, despite the fact that it is extremely rare for
orbital parallax to be measurable before an event substantially
returns to baseline.  This was attributed to the sharp deviations in
the light curve caused by a Neptune mass-ratio planet.  Yet
\citet{mb09266} say that the dominant source of the
$\Delta\chi^2=2789$ parallax signal derives from the MOA data in the
wings and that very little parallax signal comes from the perturbed
region.  However, we find from fitting the MOA data alone (and
excluding the perturbed region) that the parallax signal from these
data is only $\Delta\chi^2=205$, implying that the perturbation could
in fact play a major role in the strength of the parallax signal, as
\citet{angould01} anticipated.

It is very likely that planetary perturbations play a significant role
in the strength of the parallax signal in many other events as well.
This is likely to be a partial explanation for the fact that roughly
half of all microlens planets published to date have measured parallaxes.
Although historically, there has been a
lack of interest in where the parallax signal comes from,
 with respect to {\em Spitzer} observations, this question
is of cardinal importance.  If planetary events have more easily measured parallaxes than non-planetary events, then the sample of objects with measured parallaxes is biased.

Hence, in determining whether the event enters the sample, it is
essential to ask whether it would have a sufficiently well-measured
parallax even if there had been no planet.  This means both
eliminating follow-up observations that were triggered by the presence
of a planet and also (for this purpose) replacing the actual light
curve with a fake, point lens light curve based on the event's
parameters $(t_0,u_0,t_\e)$.  This fake light curve could then be fit
to determine the strength of the parallax signal from the point lens
event. This procedure must also be applied in cases in which the
planetary perturbation is seen from \Spitzer.

A closely related issue is that an incipient planetary anomaly might 
be misinterpreted as evidence for an approach to high magnification, 
and hence trigger an ``honest'' (i.e., seemingly non-planet-related) decision
to observe the event, either ground-based follow-up observations or
observations by {\em Spitzer}.  This occurred, for example, for ground-based
observations of the second microlensing planet, OGLE-2005-BLG-071
\citep{ob05071}.  Such observations by {\em Spitzer} might enable
a parallax measurement that would not have been possible for a point-lens
event.  This must be checked in all cases, which again can be done
through the use of fake light curves.

By contrast, it will be relatively rare for follow-up data to play a major
role in the determination of the event's point-lens parameters
simply because the main features of the event that enter
the parallax measurement usually derive from long-term 
observations and so
are well measured from survey data.  However, high-magnification
events can be an exception, primarily because dense coverage of the
peak is often required to determine the ground-based impact parameter,
particularly if the surveys cover the field at low cadence.
This was exactly the case for OGLE-2014-BLG-1049, one of the 21
events analyzed by \citet{21event}.  However, in the great majority of
cases (including this one), an improved parallax measurement 
is simply one of the benefits of
conducting follow-up observations of parallax candidates. The primary motivation is generally the increased probability of detecting planets \citep{griest98}. The only
exception would be if these critical near-peak observations were triggered
by the known presence of a planet (rather than the hope of finding one).
We expect such planet-triggered parallax-assisting 
measurements will be extremely rare and mention them 
primarily for completeness.

{\section{Principal Ingredients for Developing Selection Criteria}
\label{sec:ingredients}}

At a fundamental level, there are only two considerations in deciding
whether to choose one event over another for {\em Spitzer} observations.
First, how sensitive to planets is that event?  Second, how likely is
it that if {\em Spitzer} observations are undertaken, a microlens
parallax will actually be measured?  Thus we may schematically define
a ``quality factor'' $Q$ for the experiment
\begin{equation}
Q = \sum_i S_i P_i ,
\label{eqn:qual}
\end{equation}
where $S_i$ is the planet sensitivity of the $i$th event chosen, and
$P_i$ is the probability that it will yield a parallax measurement.
Then the goal in developing selection criteria should be to maximize
$Q$.  One issue posed by Equation~(\ref{eqn:qual}) is that planet
sensitivity is actually a function of planet properties.  However,
this is easily resolved by adding two additional indices to $S_i$
to specify these properties.  We return to this complication in 
Section~\ref{sec:quantitative}.
A more fundamental challenge is that both $S_i$ and $P_i$ may be poorly
known at the time that the decision must be made to initiate observations (c.f. Figure \ref{fig:lcs}).
The first step toward figuring out how to proceed in the face of these
uncertainties is to review what makes an event sensitive to planets
and how much about this can be known at any given stage in its 
evolution, and to then address the corresponding questions about
the measurability of its parallax.

{\section{Planet Sensitivities}
\label{sec:planet-sens}}

After an event is over, its ``sensitivity'' to 
planets\footnote{In other planet-finding contexts, e.g. radial
  velocities, ``planet sensitivity'' is often referred to as
  ``detection efficiency.''}
can
be rigorously defined as a function of two variables, the planet-star
mass ratio $q$ and the planet-star separation $s$ (in units of $\theta_\e$).
See, for example, \citet{gaudisackett}, \citet{gaudi02}, \citet{gould10},
and \citet{cassan12}.  However,
when choosing an event for additional observations while it is evolving,
one must be guided by a more qualitative understanding of what properties
make the event sensitive to planets and judge how likely it is that these
will appear.

\subsection{Ground-based Microlensing Observations}
\label{sec:ground}

Because microlensing events depend on the chance alignment of two
stars at radically different distances, for the most part, these
events cannot be predicted in advance. Hence, microlensing surveys
monitor millions of stars toward the Galactic bulge, where the stellar
density, and therefore the microlensing event rate, is highest. If the
survey team sees a star brightening in a manner consistent with
microlensing, they issue an alert announcing a new microlensing event.

Microlensing surveys monitor these fields at a variety of
cadences\\ (e.g. \url{http://www.astrouw.edu.pl/$\sim$jskowron/ogle4-BLG/}).
``High-cadence'' fields are monitored one to several times per hour
(e.g. column 2 of Figure \ref{fig:lcs}), which is sufficient to
characterize small planetary signals from terrestrial-mass
planets. ``Moderate-cadence'' fields are observed several times per
night, which can capture planetary signals from ice giants.
``Low-cadence'' fields are monitored once per night or less than once
per night (e.g. column 1 of Figure \ref{fig:lcs}) and are generally
focused on producing alerts of ongoing microlensing events that can
then be monitored more intensively by follow-up groups, although these
survey observations themselves are occasionally sufficient to
characterize large (gas-giant) planets.

In the context of this paper, we will focus on survey data from two
sources. The primary data will come from the OGLE-IV survey
\citep{ogleiv}, whose sky coverage is given in the above URL. In
addition, we will consider data from the new Korea Microlensing
Telescope Network (KMTNet) fields, which will be conducting its first
year of routine microlensing survey observations. Under the assumption
that these observations are carried out, we will include in our
evaluations data from the four core KMTNet fields, which will be
observed many times per hour.

{\subsection{Planet Sensitivity: Qualitative Features}
\label{sec:qualitative}}

Microlensing detects planets around the lens stars as
perturbations to a standard (point-lens) microlensing event.
That is, the microlensing event is overall dominated by the
gravitational potential of the lens star (host), which splits the light into
two images whose position and size change as a function of time. If a
planet interacts with one of those images, it creates a perturbation
that distorts the shape and total magnification of that image, which
can lead to a detectable signal. 
See \citet{gaudi12}, in particular Figures 4 and 5.

The total ``planet sensitivity'' of a given event depends on two
factors. The first is the intrinsic sensitivity of the event to
planets.  Larger planets, and planets that are closer to the ``Einstein
ring'' (circle with angular radius $\theta_\e$) are easiest to detect.
The larger the images are, the more sensitive they are to
planetary perturbations.  Hence, more highly magnified events are
more sensitive to planets, i.e., to smaller planets and 
to planets that are farther inside or outside the Einstein ring
\citep{gouldloeb}.  The
most extreme example would be the high magnification events 
(peak magnification of the underlying point lens event $A(t_0)=A_\max\ga 100$) in which the images form 
an almost perfect ring that probes a wide range of separations 
and is easily perturbed \citep{griest98}.

The second factor affecting planet sensitivity is how well the data
cover potential perturbations.  The quality of this coverage is
defined by two principal characteristics: cadence and photometric
precision.  Regarding the first, planetary perturbations typically last
between a few hours and a few days, so a cadence that is a factor of $\sim 10$
more frequent than that is necessary to characterize those
perturbations. Since such perturbations may occur at any time during
the event, the sensitivity will be greatest if the data are 
continuous during this period.  However, if observing resources are
limited, then restricting continuous observations to the most highly
magnified (hence most sensitive) parts of the light curve may be
the most productive approach.  Nevertheless, planets can appear in
all parts of the light curve, even after the main event is over or
before it began (e.g., \citealt{ob08092}), 
so that observations are never ``wasted''.

Photometric precision is mainly governed by source brightness.  This
factor therefore favors intrinsically bright sources, but highly
magnified sources can also be bright (just at the moment that they
are most sensitive to planets).

Even though we have not yet examined the other key determinant of $Q$
(i.e., the probability of measuring $\bpi_\e$, Section \ref{sec:parallaxprob}), we can already draw a few
general lessons from the above analysis of planet sensitivity.
First, higher peak magnification is the best single indicator for
choosing events (provided that the peak region can be intensively
monitored from the ground).  Second, for objectively chosen events,
those that are in high-cadence survey fields are substantially
more valuable than those that are not.  This is because high-cadence
events can yield planets long before the onset of {\em Spitzer} observations,
or even before the event was recognized as microlensing, whereas
low-cadence events generally cannot.  Third, for events in low-cadence
fields that achieve relatively high-magnification, it is important to 
mobilize follow-up observations prior to peak.  If these are
chosen subjectively, then the desirability of choosing this
target will depend critically on the expectation that such follow-up
observations will be carried out.

{\subsection{Planet Sensitivity: Quantitative Determination}
\label{sec:quantitative}}

The question of quantifying the planet sensitivity $S_i$ of each
event is mainly outside the scope of the present paper
because this can only be done after the event is largely over and
hence after all of the observing decisions that are the subject of this paper
have already been made.  That is, these decisions must be made on the
basis of the qualitative indicators discussed above.
Nevertheless, for completeness
we present here a broad overview of the relevant issues.

Planet sensitivity
is measured as a function of two parameters $(s,q)$ and so can be
formally written $S^{s,q}_i$.  These are two of the seven basic
parameters that are the minimum needed to describe a planetary event.
Three of the others are $(t_0,u_0,t_\e)$, i.e., the parameters of the
underlying point lens event.  The remaining two are the angle between
the source-lens trajectory and planet-star axis, $\alpha$, and the
ratio of the source radius to the Einstein radius, $\rho=\theta_*/\theta_\e$.

Historically there have been two approaches
to determining planet sensitivity.  In the first approach
\citep{rhie00}, one constructs
an ensemble of planetary light curves that vary in $\alpha$ but are
fixed in the remaining six parameters.  The values of $(t_0,u_0,t_\e)$ are
adopted from the best fit of the single-lens event.  We address the
choice of $\rho$ further below.  The remaining two parameters are just
those being tested $(s,q)$.  For each light curve, one creates fake
data points at the times of each of the real measurements, 
with values equal to those
predicted by the model and error bars equal to the those of the
real data points.  With the adopted parameters, the fit
to a fake curve without a planet would be ``perfect'', i.e., $\chi^2=0$,
so any $\chi^2$ in excess of this value must be due to the planet.  
One then fits these fake data to
a point-lens model.  Therefore, if the $\chi^2$ is above some threshold (perhaps
$\Delta\chi^2>200$ for events with moderate magnification, $A_\max<100$,
\citealt{mb11293})
then the planet is said to be detectable.  The fraction of all the $\alpha$
at fixed $(s,q)$ for which the planet is detectable is 
then said to be the sensitivity $S_i^{s,q}$.

In the other approach \citep{gaudisackett}, 
one fits the actual data with planetary models with the same
sampling of parameters, and measures the $\Delta\chi^2$ improvement
between the planetary model and the best-fit point lens model.
This method is more time-consuming but has the advantage of
simultaneously searching for all planets that may be lurking in the data
down to the adopted threshold \citep{gaudi02}.

The choice of $\rho$ is a subtle one.  For planetary events, one
often measures $\rho$ from the smearing out of the light curve
as the source passes the sharp edge of a caustic.  Hence, when constructing
fake planetary light curves, one must insert some value of $\rho$
even though this quantity is very rarely measured in point-lens events.  
The problem
is that while $\theta_*$ is usually well-determined from the color
and magnitude of the source (e.g., \citealt{ob03262}), $\theta_\e$
is not known.  We do not review the various methods used to estimate
$\rho$ in the past but simply note that for {\em Spitzer} events,
$\theta_\e = \pi_\rel/\pi_\e$ is usually known quite well because
$\pi_\e$ is measured and $\pi_\rel$ is well constrained \citep{21event},
allowing a well-constrained estimate of $\rho=\theta_*/\theta_\e$.

Current microlensing experiments have far too few detections to constrain the
full two-dimensional distribution of planets as a function of $(s,q)$.
However, for comparing to data, one can marginalize over one or both
indices, e.g.,
\begin{equation}
S_i^q = \int_{-\infty}^\infty d\log s S_i^{s,q};
\qquad
S_i^{<\rm jup} = \int_{-\infty}^{-3} d\log q S_i^q.
\label{eqn:sisq}
\end{equation}

There is one further issue that has not been previously considered in the 
literature but is quite relevant here. In the above-described procedures, it
is implicitly assumed that the observations were carried out without
reference to the presence or absence of a planet.  This has usually
been the case, and in the one notable example that it was not, the
authors took the trouble to remove the extra observations that were
triggered by the presence of the planet \citep{mb11293}.  
See also \citet{ob120406}.  However, for
{\em Spitzer} events that are chosen subjectively, a large fraction of
the nominal planet sensitivity may be due to observations before the
public announcement.  

However, as discussed in Section \ref{sec:followupsens}, there is the
possibility that a planetary perturbation in the early stages could
trigger additional observations or selection before the perturbation
is well-understood or even recognized, e.g. if the perturbation was
mistaken for a rise toward high-magnification.
To determine which hypothetical planets
should be excluded from the planet sensitivity,
we suggest that the following additional test be conducted for
each hypothetical planet that is regarded as ``detected'' based
on the full light curve: truncate the fake-data light curve 
at the decision date and fit only the points prior to this date to a
point-lens model.  We suggest that if $\Delta\chi^2<10$, then the
signal from the hypothetical planet would be too small to trigger either
follow-up or {\em Spitzer} observations.  Hence, the
hypothetical-planet detection should be accepted, but otherwise it should
be rejected.  This limit is chosen because in our experience
such  $\Delta\chi^2<10$
deviations are extremely common and so cannot possibly trigger
resource-expending actions.  However, substantially higher $\Delta\chi^2$
might well trigger an unconscious {\em Spitzer} decision announcement.

{\section{Probability of Measuring Parallax}
\label{sec:parallaxprob}}

While parallax measurements derive from a combination of ground-based
and space-based data, the limiting factor will be {\em Spitzer} data
in almost all cases.  The main reason is that the {\em Spitzer} observations of any given event are
restricted to 38 days by Sun-angle constraints\footnote{In practice, the
spread in targets over a few degrees in the Bulge allows us to stretch the
time-frame of the campaign to 40 days.}, and these 38 days
fall in an arbitrary part of the light curve.  Second, both the
cadence and quality of the data are very likely to be lower than
the ground-based data.  Finally, if the
parallax is large enough, the event as seen from {\em Spitzer} may
pass entirely outside the Einstein radius during the {\em Spitzer}
observations and so be effectively unmagnified.  
Hence, the probability that $\bpi_\e$ will be
measurable reduces in essence to probability that adequate {\em Spitzer}
observations can be obtained.
Hence, most of the discussion about ``measuring parallax'' is rooted
in the specific nature and procedures for {\it Spitzer} observations.
However, we begin by briefly discussing what it means to ``measure
parallax''.

{\subsection{Meaning of ``Parallax Measurement''}\label{sec:measurement}}

Quantities are usually said to be ``measured'' if a numerical value
can be assigned to them with some error bar and if this value is
determined to be inconsistent with zero with some specified level of
confidence, e.g., $3\,\sigma$.  For space-based microlensing
parallaxes, there are circumstances in which upper and lower limits
are sufficiently constraining, and therefore the definition of a
``measurement'' requires explicit discussion.

For example, suppose that an event is observed from the ground
with $(u_0,t_\e)_\oplus = (0.2,10\,{\rm day})$ and with $t_{0,\oplus}$
within the window of {\it Spitzer} observations, but the {\it Spitzer}
light curve is completely flat.  Also suppose that the source is
bright enough that 10\% variations in its flux would have been detected.
From these (lack of) measurements, together with the fact that
{\it Spitzer} was $\sim 1 {\rm AU}$ from Earth at the time of
observations, one could conclude that $\pi_\e > 1.5$, but no
specific value of $\pi_\e$ could be assigned.  Although not a ``measurement''
by traditional standards, this lower limit would be highly constraining.
That is, it would imply that the projected velocity  would be
$\tilde v \equiv {\rm AU}/(\pi_\e t_\e)> 115\,{\rm km\,s^{-1}}$, implying
that the lens was very likely in the near disk, $\pi_{\rm rel}>0.1\,{\rm mas}$.
Hence, if there were no planet discovered, the distance would
be statistically well enough constrained to enter the cumulative
distance distribution function.  If the event proved to have a planet,
then it is likely that $\rho$ (and so $\theta_\e$ would be measured),
which would permit a strict lower limit on $\pi_{\rm rel}=\theta_\e\pi_\e$
and a strict upper limit on $M=\theta_\e/\kappa\pi_\e$.

At the opposite extreme, if for the same $(t_0,u_0,t_\e)_\oplus$,
the {\it Spitzer} and Earth-based lightcurves
appeared identical, this would be consisent with $\pi_\e=0$, which in
traditional terms might be considered as ``no measurement''.  However,
if this consistency were quite tight, say $\pi_\e<0.01$, then the
projected velocity would be constrained $\tilde v > 17,300\,{\rm km\,s^{-1}}$,
implying that the lens certainly lies in the bulge.

Hence, the final sample must be defined as events that yield true
measurements or {\it either} upper or lower limits on $\pi_\e$ (or
both).  The exact limits cannot be defined in advance because there is
not yet enough experience with {\it Spitzer} parallax measurements to
determine what are reasonable limits.

{\subsection{{\em Spitzer} Procedure}
\label{sec:spitzobs}}

{\em Spitzer} observation sequences can only be
uploaded to the spacecraft once per week, and hence targets can only
be changed on this timescale, and furthermore, the entire week of
observations must be planned in advance. In addition, it takes
several days to prepare the observation sequence for upload to the
spacecraft even after the targets and observation sequence have been
set. The net result is that the targets and sequence are set 3--10 days
before the observations are actually carried out. See Figure~1 of
\citet{ob140124}.

In light of these considerations, and to facilitate the discussion, we
define several variables summarized in Table
\ref{tab:definitions}. First, we define $t_{j, \rm dec}$ to be 6 hours
prior to the time that observing choices must be forwarded to {\em
  Spitzer} operations for a given observing ``week'' $j$, i.e., $t_{j,
  \rm dec}=$ Monday UT 15 - 6 hrs.  Experience shows that this is the
latest time that new information can be reliably incorporated into the
observing request without risking the introduction of serious errors.
Given the day-of-the-week constraints, the $\sim 40$ day campaign, and
a start date of June 3rd, $j$ takes on values from 1 to 7.

The time of the first possible \Spitzer\, observations of a given
event is defined as $t_{\rm first}$. For simplicity, we will let
$t_{\rm first}$ be when those coordinates could first be observed by
\Spitzer, even if the event is not discovered until afterward this
date.  Finally, we define $t_{j, \rm next}$ as the time of the first
possible observation that can be requested at $t_{j, \rm dec}$, and we
define $t_{\rm fin}$ as the final possible observation of an event
before the {\em Spitzer} observing season ends (because of Sun-angle
restriction and/or the end of the allocated observations).  Note that
$t_{j, \rm dec}$ and $t_{j, \rm next}$ change each week, while $t_{\rm
  first}$ and $t_{\rm fin}$ are defined for each particular event (set
by the 38-day Sun-angle constraint). In general, for events selected
for the first week of observations, $t_{1, \rm next}=t_{\rm first}$.

{\subsection{{\em Spitzer} Observation Cycles}
\label{sec:spitzcyc}}

Generally speaking, we expect most or all of the available \Spitzer\,
time to be devoted to this program during the $\sim 40$ day observing
window. Hence, given continuous observing time, {\em Spitzer}
observations in a given week can be carried out most efficiently if
the targets are organized in concatenated ``cycles'' moving
West-to-East through the Southern bulge and then East-to-West through
the Northern bulge.  Each event can then be given a priority $n$,
which designates that it will be observed each $1/n$ cycles.  That is,
if $n=1$ it will be observed every cycle through the Bulge, and if
$n=8$ it will be observed every eighth cycle.  We expect approximately
eight cycles per day, each lasting $\sim 2.4$ hours, with the exact
number determined by the total observation time allotted and the total
number of targets per cycle. This is discussed in more practical
detail in Section \ref{sec:realloc}.

{\subsection{{\em Spitzer}'s Role in Parallax Measurements}
\label{sec:spitzrole}}

As originally conceived, the standard way to measure satellite
parallax was to observe the peak of the light curve from the satellite
and the full light curve from the ground. This requires only partial,
but very specific, light curve coverage from space. Therefore, one of
the goals of \Spitzer\, observations is to try to capture this peak in
as many cases as possible.  Although this measurement is nominally
still subject to the four-fold degeneracy in $\bpi_\e$
\citep{refsdal66,gould94}, \citet{gould95} showed that these
degeneracies could be partially or fully broken by measuring the very
small difference in $t_\e$ as seen from the two vantage points, and
this idea was then investigated in extensive simulations
\citep{boutreux96,gaudi97}. Hence, this goal of observing the peak of
the event guided the 2014 \Spitzer\, campaign.

For many years it was believed that because of the four-fold
degeneracy, parallax measurements would not be possible if the
satellite observed only the rising or falling side of the event, but
did not capture the peak. Coverage of the peak would in fact be
required if one needed to derive independent point-lens parameters
$(t_0,u_0,t_\e)$ from ground-based and space-based observations
(as is necessary in \citealt{gould95} to break the degeneracy).
  However, \citet{21event} showed that the four-fold degeneracy can
  usually be broken by a combination of the so-called ``Rich
  argument'' and kinematic priors derived from a Galactic model.  Once
  the problem of the four-fold degeneracy is removed, the requirements
  on the satellite data are drastically reduced.

First, $t_{\e,\rm sat}$ can be regarded as ``essentially known,''
so that it is only necessary to determine two satellite parameters 
($t_{0,\rm sat},u_{0,\rm sat}$) to measure the parallax.  Of course, $t_\e$
is actually slightly different as seen from Earth and the satellite
because they have a relative motion of $\sim 30\,\kms$ in the East direction.
However, the resulting difference in $t_\e$ is directly determined by
$\bpi_\e$, so while it is not strictly the case that $t_\e$
is irrelevant, it remains true that only two independent 
light curve parameters
must be derived from the satellite light curve.

Second, the source flux parameter $f_{s,\rm sat}$ for the satellite can be
determined independently of the satellite light curve using a
color-color relation derived from field stars combined with the
measured color ($I-H$ or $V-I$) derived from the ground-based light curve.
\citet{21event} obtained typical precisions for $f_{s,\rm sat}$ 
of 5\% in the cases
for which they had good $H$ or $V$ data.  However, it remains necessary
to determine $f_{b,{\rm sat}}$ from the light curve, which constitutes
a third parameter that must be derived from the {\em Spitzer} data.

Then, from simple parameter counting, it is in principle enough to measure
three non-colinear points on the light curve to measure the parallax
\citep{dong07}, e.g., one point that is ``known'' to be at baseline
and two others at different magnifications.  In practice, more points
are usually needed to have confidence in the measurement and to have
checks against discrete degeneracies.  However, it would be enough,
for example, to track the falling part of the light curve from the
time that the source exited the Einstein ring until it had dropped
by 30\% in magnification, i.e., approached baseline.

For events that are well before peak as the {\em Spitzer} window ends, the
situation is less straightforward because there would probably not
be any baseline and the short duration of the observations might
not yield any measurable change of slope.  However, such events
could be recovered by post-event baseline observations, either six months
later (when the Bulge is not visible from Earth, so ordinary satellite parallax
observations are not feasible) or the following year.  Hence, $t_{\rm fin}$
could be considered as a date in the distant future rather than the
end of the current 38-day observing window.

Therefore, there are two different channels through which parallaxes
can be measured with \Spitzer\, for point lens events. First,
\Spitzer\, can observe just the peak of the light curve. Second,
\Spitzer\, can observe either the rising or the falling side of the
event plus some measurement of the baseline. Finally, if
the event has a binary lens, features from the binary may be used to
measure the parallax. However, this situation is more complex since
binary perturbations last long enough that they may not be fully
captured by the \Spitzer\, data and so the four-fold degeneracy may
persist \citep{ob141050}.

{\subsection{{\em Spitzer} Photometric Pipeline Issues}
\label{sec:pipeline}}

The feasibility of measuring $\bpi_\e$ from a given set of {\em Spitzer}
observations obviously depends on the quality of the photometry that
can be extracted from these observations.  Remarkably, none of the
wide range of publicly available {\em Spitzer}-specific photometry
packages is well matched to the problem of time series of variable
stars in crowded fields.  As a result, the limits of what can be
achieved from such photometry are not well understood.

All the main elements required to solve this problem are at hand,
but they have not so far been combined.  First, the {\em Spitzer}
pixel response function (PRF) is extremely well understood.  That
is, if a point source has a known flux and known position relative
to the optical axis, then the response of all pixels can be predicted
to much higher precision than is relevant for the relatively faint
sources that are studied in microlensing experiments.  The positions
and fluxes of the great majority of sources in the microlensing fields
are known to be constant on the timescales of the 38-day {\em Spitzer}
observing window.  Moreover, the locations of all field sources 
that are bright enough to be relevant are known from ground-based
optical astrometry at much higher precision than is needed for
{\em Spitzer} photometry, while the location of the microlensed
source is typically known with even higher precision in this
optically based frame.  Even the approximate $3.6\,\mu$m fluxes
(other than the microlensed source) are known from optical $V/I$
photometry and fairly robust local-field $V/I/3.6\mu$m color-color
diagrams.  Hence, a conceptually straightforward procedure would
be to forward model the ensemble of $n$ images with one flux parameter
for each non-microlensed source and $n$ flux parameters for the
microlensed source. Intrinsically variable stars could be recognized
as poor fits in this process and either ignored (if they were sufficiently
far from the lensed source) or modeled with $n$ parameters instead of
just one.

We are working on such a pipeline, but since criteria for 2015 observations
(beginning in June) are required several months in advance, we must
assess likely {\em Spitzer} performance based on applying existing
pipelines to 2014 {\em Spitzer} microlens data.  These each contain
some (but not all) of the advantages of the ideal pipeline outlined
above.  For example, the MOPEX pipeline fully incorporates the PRF
but does not hold stellar positions constant, nor does it hold
the flux of field stars constant.  The well-known DoPhot \citep{dophot}
pipeline can be applied to images formed by combining the six 30s dithered
images at each epoch.  It can hold stellar positions constant but
does not incorporate any information about the PRF.  We also applied
a variant of the ISIS pipeline, which uses image subtraction to the
same combined images.  Although this pipeline normally outperforms
DoPhot for ground-based microlensing data (with some exceptions),
we find that the lack of PRF information generally affects ISIS 
more adversely than DoPhot.

We conduct a purely empirical investigation, using 47 events from the
2014 {\em Spitzer} microlensing ``pilot program'' that have enough
points to potentially construct a coherent light curve.  We consider
the photometry from MOPEX, DoPhot, and a preliminary version of our
own pipeline. We create an {\it optically-based} effective {\em
  Spitzer} (``$L$-band'') magnitude (since prior to obtaining {\em
  Spitzer} data we have no independent knowledge of the true {\em
  Spitzer} flux).  This is defined
\begin{equation}
L_\eff \equiv I - 0.93 A_I - 1.3 + 0.5\Theta(I - A_I - 17.2),
\label{eqn:optspitz}
\end{equation}
where $A_I$ is the extinction in $I$-band \citep{nataf13} and
$\Theta$ is the Heaviside step function.  We stress that no 
precise physical meaning should be attached to $L_\eff$.  It is
simply an approximate predictor of the {\em Spitzer} flux based
on optical data.  The $\Theta$ function divides all stars into
two types: turnoff stars ($\Theta=1$) and low-luminosity
giants ($\Theta=0$).  The justification for this approximation is
that significantly fainter (and redder) dwarfs generally will
not enter our sample and significantly brighter (and redder) giants
are very rare.  Of course, by limiting ourselves to two classes of stars we
are still ignoring evolution over the sub-giant branch.  However,
in the general case, it is not possible to make a finer distinction, 
particularly before a detailed investigation of an individual event has been
made, as is almost always the case when one must make the decision
about whether to monitor a particular event.

Using this proxy, we find that it is usually not possible to obtain
good photometry with existing software unless there is at
least one point with $L_\eff<15.5$.  We therefore use this
criterion as our principal guideline for deciding whether parallaxes
can be measured for particular events.  This may appear too conservative
in that there will almost certainly be photometry improvements by the
time that the data are analyzed. On the other hand, when making
decisions about {\em Spitzer} observations, one must use
the simplified ``assumption'' that {\em Spitzer} will see the same
brightness source star as it would if it were observing from Earth
because the true magnification as seen from {\em Spitzer} is unknown.
That is, the whole point of the experiment is that the {\em Spitzer}
and Earth-based light curves will differ by an intrinsically 
unpredictable amount.  In particular, the source could be less
magnified as seen from {\em Spitzer} than from the ground.
Thus, we adopt $L_\eff<15.5$ as a good balance between these two
considerations.

An additional consideration, given that the IRAC pixels are
1.2$^{\prime\prime}$, is that the target may be blended with other
stars in the crowded field, which can affect the quality of the
photometry. The severity of this blending depends both on the
separation of the blend from the target and on their relative fluxes
at $3.6\mu$m. While the separation of potential blends can generally
be determined from existing, higher-resolution, ground-based data, the
relative fluxes cannot. Furthermore, in part because of the problems
with the photometric pipelines, we have not been able to determine
exactly what criteria can be used to assess whether or not a given
blend star will cause a problem for the photometry. Hence, while we
are aware of this issue, it is not possible to account for it at the
present time.

{\section{Objective Criteria}
\label{sec:objcrit}}

\begin{figure}
\includegraphics[width=0.9\textwidth]{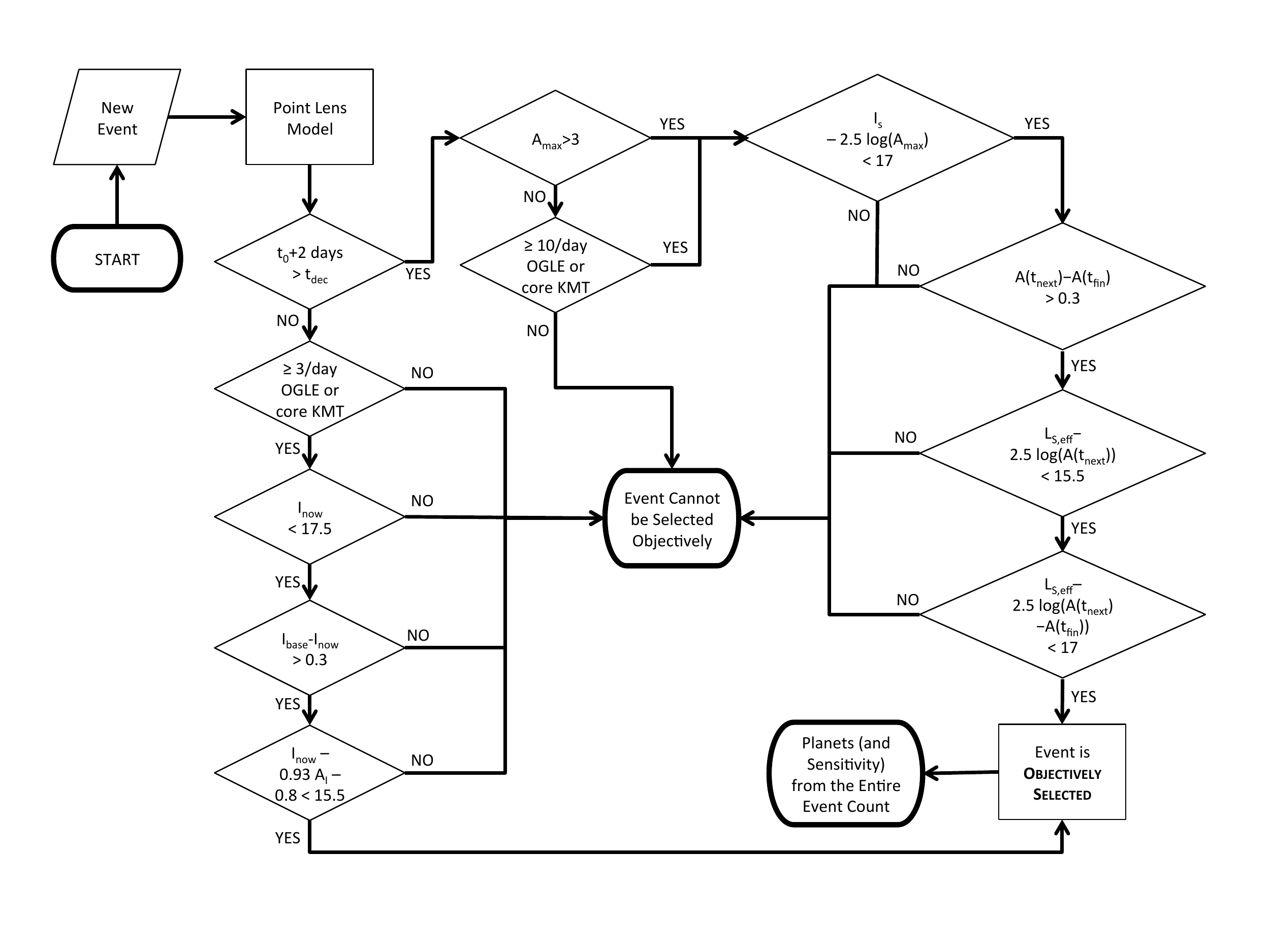}
\caption{Flowchart illustrating the process for objectively selected
  events. An event may be objectively selected either before or after
  the peak, but must meet all of the selection criteria for that
  category. \label{fig:obj}}
\end{figure}

As discussed in Section~\ref{sec:obj}, there is an extremely strong
reason for choosing as many events as possible based on purely
objective criteria: all planets (and planet sensitivity) from the
entire event can be included in the sample.  However, there is also
a huge potential for wasted {\em Spitzer} observations if these criteria are not
sufficiently restrictive. Hence, we have opted for a conservative approach.

An important point to keep in mind is that for events in low-cadence
fields, there is no major advantage to selecting the event objectively
because such events have very little sensitivity to planets in the
absence of follow-up data (Section \ref{sec:planet-sens}).  Their
sensitivity will only be substantial if higher-cadence (usually
ground-based) observations are triggered.  If this recognition also
triggers {\em Spitzer} observations (or rather, commitment to such
observations) at the same time, then essentially no planet sensitivity
is lost. Column 1 of Figure \ref{fig:lcs} demonstrates that these
low-cadence events are also extremely hard to predict.

Another point to keep in mind is
that it is substantially easier to predict the future course of
events that have already peaked than those that are still rising (compare panels 1e and 2e to earlier panels in Figure \ref{fig:lcs}),
and hence to estimate accurately whether a successful parallax measurement
can be made.  This fact is especially important for events that have
peaked before the {\em Spitzer} campaign has begun.  For events that
peak during the campaign, the probability of measuring $\bpi_\e$
can be substantially enhanced if {\em Spitzer} observations are made over peak,
i.e., before such secure information about the event parameters is available.

Guided by these considerations (and others related to subjective
selection that are discussed below), J.C.Y. and A.G. developed some
preliminary objective criteria, and then (independent of these criteria)
each individually analyzed 242
events based on OGLE and MOA data obtained up through June 3, 2013, which
is the analogous time to the first decision time in 2015.
These events had been pre-selected based on very loose criteria
from about 1000 events that had
been found by these collaborations by this date.  For each event, 
they decided whether it should be chosen for hypothetical
{\em Spitzer} observations to begin three days later, and if so
at what cadence.  All disagreements were discussed and the final
joint decisions were subsequently evaluated based on comparison to the full 2013
light curves.  The agreed-upon sample contained all nine events that
were selected by the preliminary objective criteria, and also many
that were not for a total of 44 events.  Based on this detailed analysis J.C.Y. and A.G.
refined the objective criteria for selection and the objective cadence
choice, both of which are listed immediately below, and also developed
general guidelines to subjectively choose events, which are discussed
in Section~\ref{sec:subchoice}.

Figure \ref{fig:obj} summarizes the process for objectively selecting
events. In this scheme, events may be divided into two categories:
events that have already peaked ($t=t_0+2$ days) and events ``before''
the peak ($t<t_0+2$ days). In order to be objectively selected, the
event must meet all criteria for the appropriate category. We begin by
discussing events that have peaked, because they are generally better
understood (i.e. the model fits have converged and their future
behavior is well-constrained).

\subsection{Events that have Already Peaked}
\label{sec:pastpeak}



\begin{enumerate}
\item[]{A1) $t_0 + 2\,{\rm days} <t_{j, \rm dec}$}
\item[]{A2) Either}
\begin{enumerate}
\item[]{a) in an OGLE field w/ cadence $\geq 10\,{\rm day}^{-1}$, or}
\item[]{b) in a ``core'' KMTNet field, or}
\item[]{c) $A_{\rm max}>3$ and in an OGLE field w/ cadence $= 3 \,{\rm day}^{-1}$}
\end{enumerate}
\item[]{A3) $I_{\rm S}-2.5\log(A_{\rm max})<17$}
\item[]{A4) $A(t_{j, \rm next})-A(t_{\rm fin})>0.3$}
\item[]{A5) $L_{\rm S,eff} - 2.5\log[A(t_{j, \rm next})]<15.5$}
\item[]{A6) $L_{\rm S,eff} - 2.5\log[A(t_{j, \rm next})-A(t_{\rm fin})]<17$}
\end{enumerate}

Criterion (A1) is simply a practical definition of ``post-peak''.
Criterion (A2) selects for events that have significant planet
sensitivity.  For events in high-cadence fields this is essentially
any event (provided it meets the other criteria) because planets
can be discovered in these events far out into the wings and even
at baseline.  For other, somewhat lower-cadence events, 
the criterion demands $A_\max>3$
as a minimum indicator that the {\it Spitzer} parallax observations will be
worthwhile.  However, for events in low-cadence survey fields,
it is essential to carry out ground-based follow-up observations
to gain substantial planet sensitivity.
Hence, for events in low-cadence fields, there is no triggering 
of {\it Spitzer} observations via objective criteria, and
it will therefore be necessary to subjectively select these
events prior to peak (which may be before the {\it Spitzer} observing
window).

Criterion (A3) demands that the peak flux from the magnified source
(not including any blended light) be $I<17$. It increases the
likelihood that the ground-based photometric precision will enable
good planet sensitivity.  To date, the overwhelming majority of
planetary microlensing events have peaked $I<17$.  In addition, this
criterion is important not only to secure an accurate fit to the light
curve but to permit application of color-color relations to determine
the {\em Spitzer} source flux, both of which affect the final parallax
measurement.

Criterion (A4) is driven by the fact that parallax measurements
require a well-measured change in magnification as seen from 
{\em Spitzer}.   In practice, this means both measuring 
a flux change and independently determining the {\em Spitzer}
source flux.  For observations that begin past peak, it is
impossible to reliably fit the {\em Spitzer} light curve
for the source flux, so it must
be determined from color-color relations, which 
can be done reliably to about 5\%.  Hence, we require at least a 0.3
change in magnification (i.e., 6-fold larger than 5\%)
based on our estimate of what
will be required for reliable parallax measurements (Section \ref{sec:spitzrole}).
This criterion is equivalent to demanding that the source is still
in the Einstein ring (at the time of the next possible observation)
for the case that the final possible observation is well outside
the Einstein ring.  

Criterion (A5) derives directly from the difficulty of extracting {\em Spitzer}
photometry unless at least one point is brighter than $L_\eff<15.5$ (Section \ref{sec:pipeline}).
Criterion (A6) demands a minimal flux change in {\em Spitzer} flux.
It will be automatically satisfied for the great majority of stars
that satisfy criterion (A5) and is included to guard against including
(at least automatically) events that are not predicted to change much
over the remainder of the observations.  As noted in the justification
for criterion (A3), measurable flux changes are crucial for parallax
measurement.  Criterion (A6) ensures that not only the magnification
changes, but the flux itself changes by a significant amount.

{\subsection{Rising Events}\label{sec:rising}}

\begin{enumerate}
\item[]{B1) $t_0 > t_{\rm dec}-2\,$days}
\item[]{B2) in OGLE field w/ cadence $\geq 3\,{\rm day}^{-1}$ or in a ``core'' KMTNet field}
\item[]{B3) $I_{\rm now} < 17.5$}
\item[]{B4) $I_{\rm base}-I_{\rm now} > 0.3$}
\item[]{B5) $L_{\rm eff, dwarf, now} = I_{\rm now}-0.93A_I-0.8 < 15.5$}
\end{enumerate}

Criterion (B1) is just the practical definition of a rising 
event\footnote{Technically, this also includes events $<2\,$days past peak.
The reason for this choice is that it can be difficult to be confident
that an event has indeed peaked unless there are data after the peak to 
demonstrate this explicitly.} (i.e.,
the complement of criterion (A1).

Criterion (B2) restricts this entire class of objective selection to
fields with moderate-to-high cadence, $\geq 3\,{\rm day}^{-1}$,
to ensure that the selected events have good sensitivity to planets
over the entire light curve. The primary concern is that once an event
has met the full set of rather restrictive criteria, it will be close
to the peak and a large fraction of the event will be over. As
discussed in Section \ref{sec:planet-sens}, without substantial
follow-up data, low-cadence events have very little practical planet
sensitivity. Hence, it is not worthwhile to observe them with
\Spitzer\, to try to measure a parallax because they will add very
little to the final analysis of the planet occurrence rate as a
function of Galactic distance. We note that most high-quality events
that fail this criterion can be selected subjectively
(Section~\ref{sec:subchoice}) well before peak with the specific goal
of triggering additional follow-up observations to improve the planet
sensitivity.

The remaining three criteria make no reference to a light curve model
and instead rely on purely empirical observables.  Again, this is
because experience shows that such models are not reliable for pre-peak
light curves.  All three make reference to ``$I_{\rm now}$'' which is
the last measured OGLE point.

Criterion (B3) assures that the event will be bright enough for
accurate measurements (necessary for both planet sensitivity and
parallax). Criterion (B4) assures that even if the source is not
blended, the event has risen at least 32\% above baseline, i.e., the
source must be (nearly) inside the Einstein ring. The combination of
(B3) and (B4) ensure that even if the source turns out to be heavily
blended, then at least the magnified flux will change significantly
compared to its present value relative to a future baseline
measurement, which will enable a measurement of the parallax.
 Finally, criterion (B5)
attempts to assure that there will be at least one point above the
photometric threshold for measuring a {\em Spitzer} light curve (Section \ref{sec:pipeline}).  This is only
``attempted'' (rather than guaranteed) since
the event may not be as magnified from {\em Spitzer}'s
vantage as from Earth.  Because the source magnitude is most likely
not known at the time of this algorithmic selection, we conservatively
assume it is a dwarf (i.e., relatively blue and so fainter as seen by
{\em Spitzer} for fixed $I$-band brightness).

\subsection{Objectively Determined Cadences}

As we have discussed, events that are selected objectively must have
objectively determined cadences.
In practice, cadences are actually defined by ``priorities'', where
priority $n$ means that the event is observed during $1/n$ of the
cycles through the microlensing fields (Section~\ref{sec:spitzcyc}).
However, we state these here
in terms of cadences, since there is a clear-cut conversion from one
to the other once the target sample is selected.  We designate the
following algorithm for setting the observation cadence for Week $j$

\begin{enumerate}
\item[]{C1) Default cadence: $1\,{\rm day}^{-1}$}
\item[]{C2) $2\,{\rm day}^{-1}$ provided that all of the following are true}
\begin{enumerate}
\item{This is first {\em Spitzer} observation period of the event, and}
\item{$t_{j, \rm next}>t_0$, and}
\item{$A(t_{j, \rm next})<1.35$}
\end{enumerate}
\item[]{C3) $2\,{\rm day}^{-1}$ for events beginning the when
less than two full weeks remain}
\item[]{C4) Stop observing the event, provided that all of the following are true}
\begin{enumerate}
\item{At least 2 weeks of objectively-determined observations are complete, and}
\item{$t_{j, \rm next} > t_0+t_\e$}
\item{$t_{j+1, \rm next} > t_0 + 2 t_\e$}
\end{enumerate}
\item[]{C5) Stop observing the event, if either}
\begin{enumerate}
\item{The 
parallax of an event has been measured from {\em Spitzer} data
already collected, or}
\item{The {\em Spitzer} light curve has already reached
baseline (so no more parallax information could be extracted from
additional observations)}
\end{enumerate}
\end{enumerate}

Criterion (C1) has been shown by \citet{21event} to be generally
adequate to make parallax measurements.  However, for events that
are leaving the Einstein ring as the {\em Spitzer} observations
begin (C2) or for which there is only a short rising observational
sequence at the end of the {\em Spitzer} window (C3), the cadence
is doubled.  These events have significantly more restricted light curve
coverage than the typical events analyzed by \citet{21event} and therefore
require higher cadence to obtain more points (so higher signal-to-noise ratio)
while the event is still significantly magnified.
(C4) imposes a reasonably conservative criterion
for halting observations.  In principle, this may cause {\em Spitzer}
to miss a key portion of the light curve because it can in principle peak either earlier or
later as seen from {\em Spitzer} than the Earth-based light curve
would predict.  However, because bulge lenses have small microlens
parallaxes, the light curve peaks as seen from Earth and {\em Spitzer}
are very close in time.  On the other hand, for disk lenses,
the {\em Spitzer} peak is usually earlier (or not much later) than
from the ground because these disk lenses tend to move in the direction
of Galactic rotation, i.e., about $30^\circ$ East of North, whereas
{\em Spitzer} is roughly due West of Earth.  See Figure 2 of \citet{21event}.
Finally, (C5) provides more specific conditions for halting
observations if the {\em Spitzer} data can be reduced and analyzed in
real-time.

{\subsection{Binary Events} \label{sec:binaries}}

For completeness, we also specify the objective selection criteria for
binary events. Unlike planetary events, binary events show
prominent anomalies that modify the single-lens light curve
significantly. Therefore, most binary events can be recognized in
advance, and the inability to model them with single-lens light curves
makes all selection criteria based on single-lens modeling (i.e.,
Sections \ref{sec:pastpeak} and \ref{sec:rising}) fail in most
cases. As a consequence, in the 2014 pilot program we subjectively
selected binary events, such as OGLE-2014-BLG-1050 \citep{ob141050},
for observations because their nature as binary events had been
confirmed.

Therefore, in order to enable statistical studies of stellar binaries
one has to have objective selection criteria. After reviewing those
binary events from the 2014 season, we decide to use the following
criteria and cadence:

\begin{enumerate}
  \item{Begin \Spitzer\, observations if}
  \begin{enumerate}
    \item{The ground-based light curve is in a U-shaped trough,}
    and
    \item{$L_{\rm eff, trough} < 16$.}
  \end{enumerate}
  \item{End \Spitzer\, observations}
  \begin{enumerate}
    \item{Either:}
    \begin{enumerate}
      \item{One full week after the {\it Spitzer} light curve exits the caustic,}
      or
      \item{Both:}
      \begin{enumerate}
        \item{One full week has passed since the ground-based light curve exits the caustic,}
        and
        \item{The {\it Spitzer} light curve is shown never to have entered the caustic.}
      \end{enumerate}
    \end{enumerate}
  \end{enumerate}
  \item{Default cadence of 1 day$^{-1}$.}
\end{enumerate}

{\section{Guidelines for Subjectively Chosen Events}
\label{sec:subchoice}}

\begin{figure}
\includegraphics[width=0.9\textwidth]{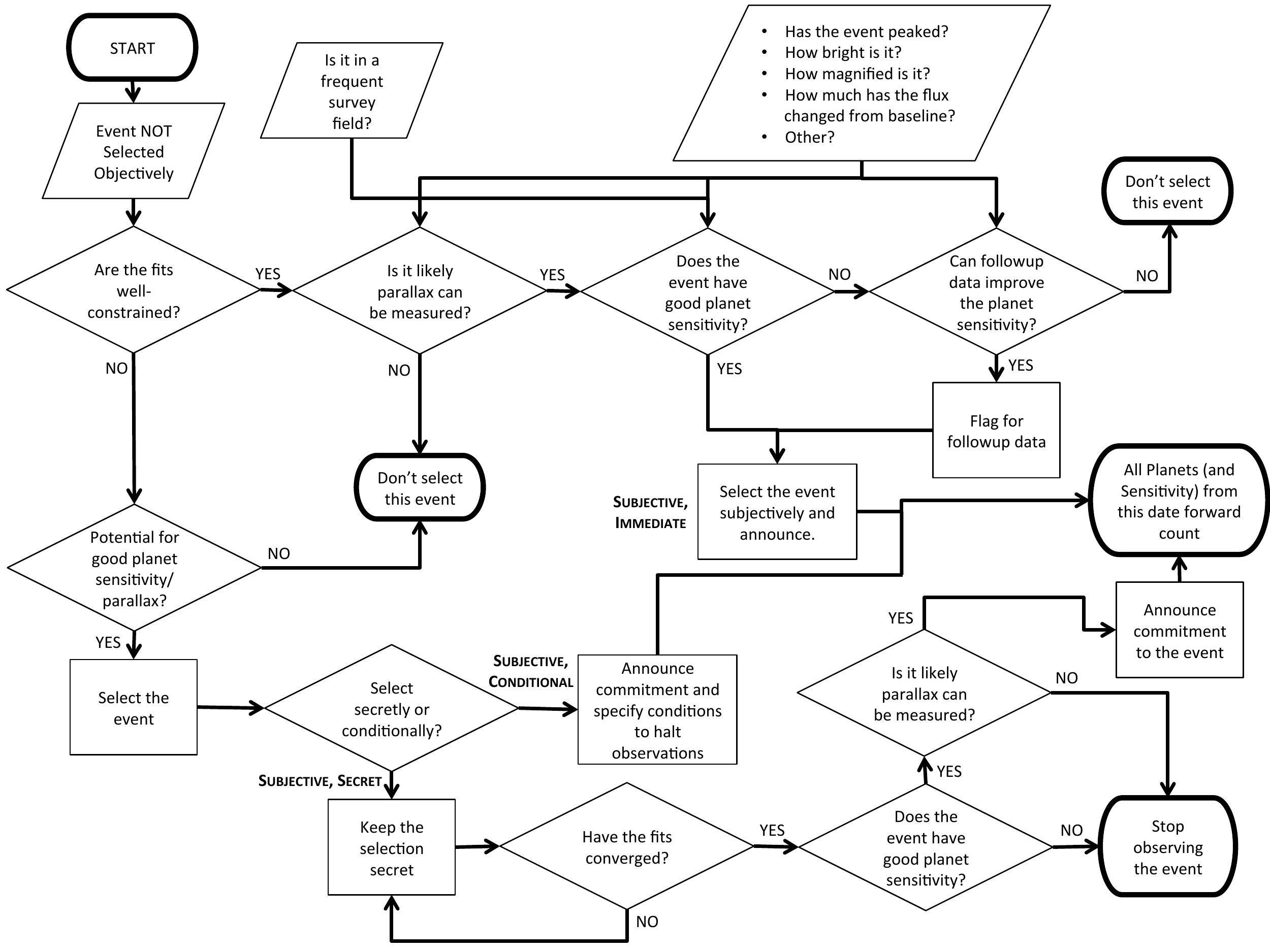}
\caption{Flowchart illustrating the process for subjectively selecting
  an event. By definition, this is less quantitative than the process
  for objectively selecting an event (see Figure \ref{fig:obj}), but
  the underlying considerations are the same in both cases, namely:
  ``Does the event have good planet sensitivity?'' and ``Is this event
  likely to yield a \Spitzer\, parallax?''.\label{fig:sub}}
\end{figure}

We have already discussed in Section \ref{sec:sub} the various reasons
one might want to subjectively select an event and how that might
affect the type of subjective selection (Table
\ref{tab:selection}). Figure \ref{fig:sub} summarizes the decision
process that can lead to subjective selection of an event.

The most straightforward type of subjective selection is ``Subjective,
immediate'' in which an event is immediately selected for
observations, committed to, and has its cadence specified. Such a
decision may be made at any time, including before the start of
\Spitzer\, observations. There are two primary types of events that
might be selected this way. First, events that are discovered and well
understood before $t_{j, \rm dec}$ but might not meet the objective
criteria. Committing to observations immediately allows more of the
planet sensitivity of that event to be captured. The second type of
event is a well-understood, low-cadence event that requires follow-up
observations in order to capture significant planet sensitivity.

However, we expect that the great majority of events will be selected
under the categories ``Subjective, conditional'' or ``Subjective, secret'' because at time $t_{j, \rm dec}$
there is no way to distinguish between
events that will reach moderate or higher magnification ($A_\max>3$)
sometime before the start of the next cycle of {\em Spitzer} observations,
$t_{j+1, \rm next}$ 
(i.e., $t_{j, \rm dec} + 10\,{\rm day}$), and events that will turn over
at low magnification and low flux levels (e.g. compare the two fits in panel 1b of Figure \ref{fig:lcs}).  Moreover, even among those
that are likely to achieve satisfactory magnification, it cannot
be decided automatically whether suitable follow-up resources can be 
allocated to a specific event, given a wide variety of operational constraints.
And finally, it may be impossible to determine which of these
events will get to high or very high magnification based on routine
survey data.  However, additional investigation, including additional
follow-up data and/or color information from survey and/or follow-up
observations may resolve this question.

As we have already discussed above, subjective
decisions may be made before or after $t_{\rm first}$, but they
must specify the cadence (or cadence algorithm) at the time of
commitment.  To be robust, any algorithm must be based on readily
available data, such as the $I$-band light curve and the $I$-band
field extinction.

Given the above factors, together with the fact that planet sensitivity
is heavily skewed toward higher-magnification events, it is inevitable
that {\em Spitzer} observations will be triggered for a large number
of events of uncertain prospects, and therefore that the majority of these
must be terminated promptly after the event fails to rise to the
level that permits significant planet sensitivity 
and also permits its parallax to be measured.
Otherwise, a substantial fraction of observing time will be wasted on
useless events.

Hence, the principal issue will be deciding how to frame this failure
in either of the two cases (``Subjective, conditional'' or ``Subjective, secret'').
This will have to be done on a case-by-case basis at the time the
events are subjectively chosen, since the uncertain nature of such
events makes it impossible to develop strict guidelines in advance.
In the first case, this framework must provide a 
guide to terminating observations (e.g., ``stop observations
next week if event does not reach $I<16.2$ by the next decision point, $t_{j+1,\rm dec}$'').
In the second, it must provide a guide as to whether
or not the team should commit to observations of the event.

To the extent possible, the criteria governing such terminations (or
continued observations) should be framed in terms of post-peak observables
because these are more robust.  They should then be formulated as
proxies for the criteria (A1)--(A6), but with
$t_{j, \rm next}\rightarrow t_0$.  However, this implicitly assumes
that $t_{j, \rm dec}<t_0<t_{j+1,\rm dec}$ (essentially criterion
(A1)). It will in general be necessary to specify what should be done
if the event does not ultimately satisfy that criterion.
For example, it could be stated that in this case, the observations
continue at the same cadence and the decision is made during the
next week.

{\section{Objective Allocation of Remaining Observing Time}
\label{sec:realloc}}

\subsection{General Considerations} 
An important aim of the {\em Spitzer} observations, as outlined in the
proposal \citep{gouldcareyyee14} is to detect and characterize planets
from {\em Spitzer} observations themselves. Because planet sensitivity
scales with magnification, this can best
be done by monitoring higher-magnification events more intensively
from {\em Spitzer}, in particular events that are at higher magnification
as seen by {\em Spitzer}.  Hence, after the fundamental goal is
met by allocating enough observation time to reliably measure parallaxes
(as outlined in Section~\ref{sec:objcrit} and \ref{sec:subchoice})
and also to obtaining parallaxes of microlens binaries (which was also
part of the proposal but is not the subject of the present study),
the remaining time should be allocated to this purpose.  

Because these additional planet-finding observations involve allocation 
of additional time to the {\it same} set of events that are the
object of microlens parallax measurements, it is important to
isolate the decisions about this allocation from the prospect
of improving parallax measurements.  Otherwise, events with known
planets could receive additional measurements aimed at measuring
parallax, making them more likely to have good parallax measurements
relative to those that failed to show any planets. And since only
events with good parallax measurements can be included in the sample
to measure the Galactic distribution of planets, we must eliminate the
potential for bias.

At the same time, it is difficult to develop completely objective
criteria for these allocations because of the wide range of the
possible quantities of available time and wide range of the possible quality
of events to which they might be applied. 
In addition, the amount of additional time will
depend on the precise number of events being monitored. Finally, the
targets for a given week will have a wide range of magnifications,
whose distribution cannot be predicted in advance.

\subsection{Practical Execution}

We propose the following algorithm to effectively separate these
decisions while leaving adequate freedom to respond to potential
planet sensitivity by obtaining additional {\em Spitzer} observations.

Recall from Section~\ref{sec:spitzcyc} that the most efficient way to
observe these events is to cycle through the Bulge West-to-East and
East-to-West. The priority of an event sets how often an event will be
observed. Specifically, priority $n=1$ is observed every cycle whereas
priority $n=8$ is observed every eighth cycle. The exact number of
cycles depends on the the number of events, their priorities, and the
amount of time available.

All events that have been chosen for {\em Spitzer} observations,
{\it except newly selected subjective events} are rank ordered
according to the $2\,\sigma$ lower limit of their highest magnification ($A_{j,\max}$)
in the observing window [$t_{j,\rm next},t_{j+1,\rm next}$].
These are then assigned priorities that map monotonically to this
ranking.  
The break points
in this mapping are decided manually.  For example, 
$A_{j,\max}>20\Rightarrow n=1$,
$10<A_{j,\max}\leq 20\Rightarrow n=2$,
$5<A_{j,\max}\leq 10\Rightarrow n=3$,
$3<A_{j,\max}\leq 5\Rightarrow n=4$. 
The only other rule is that the priorities and the number of events in
each category result in a total number of observations that equals the
number of observations available. Hence, there are likely many choices
of break points and priorities that fulfill these two criteria
(monotonic mapping and total observations) for any given rank-ordered
set of events and amount of observing time. The final choice, which
will set the number of cycles and their duration, is at the discretion
of the \Spitzer\, team.

While the manual decision on break points might seem to allow
skewing (conscious or unconscious) of the strength of different
parallax measurements, in fact this is virtually
impossible.  The events whose parallax is poorly measured with existing data, and therefore
might require ``saving'' with additional observations, will be
at low magnification and hence cannot be helped by any manual decision
that is constrained by the monotonic mapping described above.

Here, the default meaning of ``highest magnification'' is highest
as predicted for observations from the ground.  However, it may
be possible to download and process {\em Spitzer} data sufficiently
quickly to make predictions about the course of the event as
seen from {\em Spitzer}.  In this case, the {\em Spitzer} magnifications
should take precedence.

Why do we exclude the subjectively chosen events whose observations
are just starting?  There are two reasons.  First, the future course
of these events is often very poorly understood, so the $2\,\sigma$ lower
limit on the magnification is likely to be very low and hence unlikely
to trigger the additional observations being considered here.  
More fundamentally, the cadence of these
first-week observations is subjectively decided, so no objective
procedure is required to allocate additional observations 
to planet hunting in these events if the team decides that is necessary.

The cadence (or algorithm for determining the cadence) of {\it
  Spitzer} observations must be specified at the time of the
commitment to observe a subjectively selected event. However, in practice these
cadences will almost always be set at $t_{\rm sel}$, since even for ``subjective, secret'' events $t_{\rm com}$ will generally happen before the next decision point when the cadence could be changed ($t_{j+1,\rm dec}$). In 
subsequent weeks, any change in the cadence of these events must be
through the process described above.

{\section{Conclusions}
\label{sec:conclude}}

We have outlined an approach for maximizing the planet sensitivity of
space-based microlens parallax surveys, for the particular case that
the satellite targets are chosen based on ground-based identification
of events.  This applies to current observations by {\em Spitzer}
and any narrow field-of-view, targeted observations from space\footnote{ 
It does not apply to a microlens survey by {\it Kepler} (K2) because the
field will be a large, pre-selected region rather than having the observations
targeted at specific, known microlensing events.}.

The basic principles are:

First, objective criteria are quite easy to establish for events that
have already peaked because their fits are well constrained.  
Because the criteria are objective, the entire
time span of those events, with respect to 
both planets and planet sensitivity, can be
included in the analysis.  This includes, in particular, all of the
time before peak, which lies before the onset of {\em Spitzer} observations
and also before the event was even recognized as microlensing.

Second, it is also possible to establish objective criteria for a
subset of pre-peak events. However, because these events are pre-peak,
their model fits may not be reliable, so it is necessary to define
these criteria in terms of observables.

Third, for objectively chosen events, cadences must also be determined
by objective criteria.  

Fourth, for those events in low-cadence survey fields, it is less
important to define objective criteria, because the events have low
sensitivity to planets unless additional follow-up observations are
obtained. 

Fifth, the remaining events, including those in low-cadence fields,
can be chosen subjectively, but the full cadence (or prescription for
determining the cadence) must be specified at the time that they are
selected. Subjective selection can take several forms, but the most
important aspect is when a commitment is made (and announced) to
observe the event.  It is this date that determines what planet
sensitivity and planet detections are included in the analysis.

Sixth, in the case of events that were previously chosen on subjective
grounds and that subsequently meet the objective criteria, their
objective status must take precedence in evaluating the event as part
of the sample. This assumes that a parallax measurement proved
possible based only on the (more restricted subset of) objectively
required {\em Spitzer} observations.  If not, and if the full set of
{\em Spitzer} observations yields a parallax measurement, then they
revert to subjective status.

This paper constitutes a public announcement of our objective criteria
and procedures.  If there are any updates to these, they will be posted
on arXiv as a revision to or update of this paper.

\acknowledgments

Work by JCY, AG, and SC was supported by JPL grant 1500811.  Work by JCY was
performed under contract with the California Institute of Technology
(Caltech)/Jet Propulsion Laboratory (JPL) funded by NASA through the
Sagan Fellowship Program executed by the NASA Exoplanet Science
Institute. DMN was supported by the Australian Research Council grant FL110100012.

\end{document}